%% file: main.tex
\documentclass[sigconf,screen]{acmart}

\renewcommand\footnotetextcopyrightpermission[1]{} 
\usepackage{enumitem}
\usepackage{pifont}
\usepackage{multirow,makecell}
\usepackage[skins]{tcolorbox}
\usepackage{color,xcolor}
\usepackage{listings,amsfonts}
\usepackage{caption}
\usepackage{subcaption}
\usepackage{threeparttable}
\usepackage{bbding}
\usepackage{graphicx}
\usepackage{booktabs} 
\usepackage{longtable}
\usepackage{flushend}
\usepackage[linesnumbered,ruled,vlined]{algorithm2e}
\usepackage{hyperref}
\usepackage{amsmath}
\usepackage{array}
\usepackage{algpseudocode}
\usepackage{colortbl}

\usepackage{etoolbox} \makeatletter \patchcmd{\@autoref}{\begingroup}\addto\extrasenglish{%
}

\AtEndPreamble{
	\usepackage{hyperref}

}

\newtheorem{definition}{Definition}

\begin{document}

\title{Unveiling Large Language Model Supply Chain: Structure, Domain, and Vulnerabilities}

\author{Yanzhe Hu}
\authornote{Both authors contributed equally to this research.}
\email{yanzhehu@hust.edu.cn}
\orcid{0009-0009-5318-3043}
\affiliation{%
  \institution{Huazhong University of Science and Technology}
  \city{Wuhan}
  \state{Hubei}
  \country{China}
}

\author{Shenao Wang}
\authornotemark[1]
\email{shenaowang@hust.edu.cn}
\orcid{0000-0003-3818-3343}
\affiliation{%
  \institution{Huazhong University of Science and Technology}
  \city{Wuhan}
  \state{Hubei}
  \country{China}
}

\author{Yuhan Tang}
\email{yuhantang@hust.edu.cn}
\orcid{0009-0008-3879-0083}
\affiliation{%
  \institution{Huazhong University of Science and Technology}
  \city{Wuhan}
  \state{Hubei}
  \country{China}
}

\author{Tianyuan Nie}
\email{tianyuannie@hust.edu.cn}
\orcid{0009-0003-3863-6880}
\affiliation{%
  \institution{Huazhong University of Science and Technology}
  \city{Wuhan}
  \state{Hubei}
  \country{China}
}

\author{Yanjie Zhao}
\email{yanjie_zhao@hust.edu.cn}
\orcid{0000-0001-8793-5367}
\affiliation{%
  \institution{Huazhong University of Science and Technology}
  \city{Wuhan}
  \state{Hubei}
  \country{China}
}

\author{Haoyu Wang}
\authornote{Corresponding author.}
\email{haoyuwang@hust.edu.cn}
\orcid{0000-0003-1100-8633}
\affiliation{%
  \institution{Huazhong University of Science and Technology}
  \city{Wuhan}
  \state{Hubei}
  \country{China}
}

\renewcommand{\shortauthors}{Yanzhe Hu et al.}

\input{Chapters/0.abstract}

\maketitle

\input{Chapters/1.introduction}
\input{Chapters/2.dataset}
\input{Chapters/3.RQ1}
\input{Chapters/4.RQ2}
\input{Chapters/5.RQ3}

\input{Chapters/6.discussion}
\input{Chapters/7.literature}
\input{Chapters/8.conclusion}

\balance
\bibliographystyle{ACM-Reference-Format}
\bibliography{reference}

\end{document}

%% file: Chapters/0.abstract.tex
\begin{abstract}
Large Language Models (LLMs) have revolutionized artificial intelligence (AI), driving breakthroughs in natural language understanding, text generation, and autonomous systems. However, the rapid growth of LLMs presents significant challenges in the security and reliability of the Large Language Model Supply Chain (LLMSC), a complex network of open-source components, libraries, and tools essential for LLM development and deployment. Despite its critical importance, the LLMSC remains underexplored, particularly regarding its structural characteristics, domain composition, and security vulnerabilities.
To address this gap, we conduct the first empirical study of the LLMSC, analyzing a curated dataset of open-source packages from PyPI and NPM across 14 functional domains. We construct a directed dependency graph comprising 13,486 nodes, 28,704 edges, and 180 unique vulnerabilities to investigate the structural characteristics of the LLMSC and analyze how security risks propagate through its dependency network. Our findings reveal that the LLMSC exhibits a ``locally dense, globally sparse'' topology, with 72.38\% of dependency trees containing fewer than 5 nodes, while a few large trees dominate the ecosystem, accounting for 77.66\% of all nodes. The graph is characterized by high-degree hubs, with the top 5 most connected nodes averaging 1,207 dependents each. Security analysis shows that critical vulnerabilities propagate to an average of 142.1 nodes at the second layer of dependency trees and peak at 237.8 affected nodes at the third layer. Notably, cascading risks are concentrated in critical hub nodes such as \texttt{transformers}, which directly or indirectly affect over 1,300 downstream packages.
These findings provide quantitative insights into the structural and security dynamics of the LLMSC and emphasize the need for targeted mitigation strategies to enhance ecosystem resilience. 
\end{abstract}

%% file: Chapters/1.introduction.tex
\section{Introduction}

Recently, Large Language Models (LLMs) have revolutionized artificial intelligence (AI), driving breakthroughs in natural language understanding~\cite{daye2024llmcodeunderstanding,lu2024visionunderstanding,zhao2024llmsurvey}, text generation~\cite{zhao2024llmsurvey,roberto2023generativeai}, software engineering~\cite{hou2024llm4se,jin2024agents4se,wang2024agentsinse}, and autonomous systems~\cite{wu2023autogen,wang2024ala}. However, their rapid development raises significant challenges in the security and reliability of the \textbf{Large Language Model Supply Chain (LLMSC)}~\cite{wang2024llmsc,huang2024llmsc,hu2024llmsc}. Vulnerabilities in the LLMSC compromise confidentiality, integrity, and availability, leading to risks such as biased outputs, security breaches, or system failures~\cite{owasp}. These risks highlight the urgent need to secure the LLMSC.

As defined in previous studies~\cite{wang2024llmsc,hu2024llmsc,huang2024llmsc}, the LLMSC encompasses various components critical to the lifecycle of LLMs. Broadly, it includes hardware resources, datasets, third-party components, pre-trained models, and adaptable prompt templates. 
While existing studies have primarily focused on the vulnerabilities associated with datasets and models, such as poisoning attacks~\cite{zhao2024malhug,chen2024agentpoison,zhang2024ragpoisoning} and the risks introduced by reusing pre-trained models~\cite{jiang2022ptm,jiang2023huggingface}, our work shifts the focus to the open-source components within the LLMSC. Vulnerabilities in these components, compared to those in datasets or models, are more likely to directly compromise system security, posing risks such as data breaches, unauthorized access, or system failures. Specifically, we adopt a narrower definition of the LLMSC\footnote{Also referred to as the LLM toolchain in some prior studies}, which refers to the supply chain formed by open-source components, libraries, and tools involved in developing, testing, deploying, and maintaining LLMs.
For instance, vulnerabilities in widely-used libraries such as \texttt{llama-index} and \texttt{transformers} have allowed attackers to execute arbitrary code remotely~\cite{llamaindexcve,transformerscve}, demonstrating how a single flaw can propagate throughout the ecosystem and compromise downstream applications~\cite{owasp}.

Despite the critical importance of securing the LLMSC, research on this topic remains limited. To date, the LLMSC remains largely a black box, with many aspects of its structure, dependencies, and security dynamics still poorly understood.
First, while existing studies have explored the supply chains of deep learning frameworks~\cite{tan2022dlsc,gao2024dlsc}, the LLMSC introduces unique challenges and distinctions. Traditional deep learning supply chains are often centered around a small number of core frameworks, such as TensorFlow~\cite{tensorflow} or PyTorch~\cite{pytorch}, which serve as foundational components for most applications. In contrast, the LLMSC is far more heterogeneous, with no clear consensus on which components are most critical. 
Second, while substantial research has been conducted on software supply chains in ecosystems like C++~\cite{tang2023csc}, NPM~\cite{liu2022npmsc}, or Maven Central~\cite{wu2023mavensc}, the LLMSC represents a specialized domain with its own unique characteristics. Unlike general-purpose software ecosystems, the LLMSC often includes components that are highly tailored to LLM development, such as dataset curation tools, LLM-specific deployment frameworks, and Retrieval-Augmented Generation~(RAG) frameworks. These specialized components may exhibit unique structural or functional dependencies, offering opportunities to extract insights that are distinct to the LLMSC.
Finally, as an emerging and rapidly expanding ecosystem, the LLMSC is highly dynamic. The number of open-source components relevant to LLM development is growing at an unprecedented pace, with new tools, libraries, and frameworks emerging regularly. This dynamic growth means that the LLMSC is not a static entity but a constantly evolving network of dependencies.

To bridge this knowledge gap, we conducted the first empirical study of packages in the LLMSC to better understand its structure, components, and vulnerabilities.  
Our study began with a fully automated pipeline to collect and filter open-source packages from PyPI and NPM. Through keyword-based searches and heuristic filtering, we gathered metadata and README files for 54,393 packages and curated a dataset of 13,486 LLM-related packages across 14 functional domains.  
To analyze the relationships among these packages, we constructed a directed graph where nodes represent packages and edges denote explicit dependency relationships. This graph, comprising 13,486 nodes and 28,704 edges, captures the intricate structure of the LLMSC and provides a foundation for exploring its dependency network.  
Additionally, we examined the security landscape by aggregating vulnerability data from multiple sources, including MITRE CVEs, GitHub Advisory, and huntr. After deduplication and filtering, we identified 180 unique vulnerabilities directly affecting LLMSC components. These were integrated into the graph as vulnerability nodes and affect relationships, enabling us to study how security risks propagate through the ecosystem.  
We present three Research Questions~(RQs), and the main findings are as follows.

\begin{itemize}[leftmargin=15pt]
    \item \textbf{RQ1: (Structure)} \textit{What are the structural characteristics of the LLMSC?} 
    \item \textbf{RQ2: (Domain Distribution)} \textit{What is the domain composition of the LLMSC, and how has it evolved over time?} 
    \item \textbf{RQ3: (Security Risks)} \textit{What are the security risks within the LLMSC, and how do vulnerabilities propagate?} 
\end{itemize}

\noindent \textbf{Contributions.} We make the following contributions:
\begin{itemize}[leftmargin=15pt]
    \item To the best of our knowledge, we conduct the first systematic study of the LLMSC, uncovering its hierarchical structure, domain distribution, and security vulnerabilities, thereby providing a foundational understanding of this emerging ecosystem. 
    \item We develop a methodology for constructing the LLMSC and assemble a dataset comprising metadata for 13,486 packages and 180 vulnerability reports. The dataset will be made publicly available upon the acceptance of this paper, enabling reproducibility and fostering future research. 
    \item Our findings reveal that the LLMSC is dominated by high-degree hubs, exhibits a ``locally dense, globally sparse'' topology with diamond and triangle substructures, and faces significant security risks due to cascading vulnerabilities, primarily during the first three transmissions in dependency trees.
\end{itemize}

In the remaining sections of this paper, \autoref{sec:preparation} describes how we constructed the dataset and built the LLMSC. \autoref{sec:rq1} to \autoref{sec:rq3} present the methodology and results for each RQ. \autoref{sec:discussion} discusses threats to validity. \autoref{sec:relatedworks} presents related work, and \autoref{sec:conclusion} concludes the paper.

%% file: Chapters/2.dataset.tex
\section{Dataset and LLMSC Construction}
\label{sec:preparation}
\subsection{Dataset Collection and Pre-process}
\label{sec:dataset}
\begin{figure}[!htbp]
    \centering
    \includegraphics[width=0.95\linewidth]{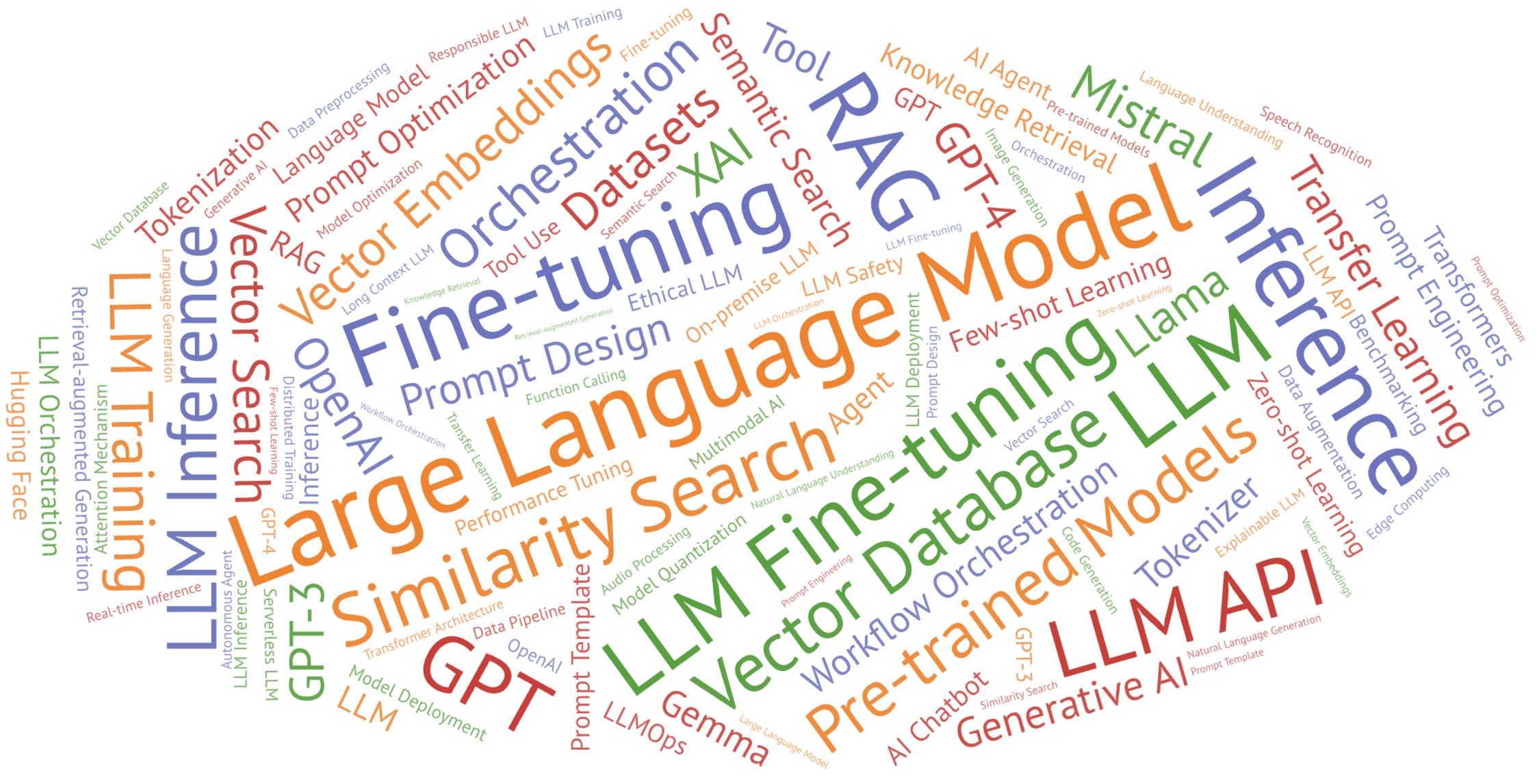}
    \caption{The word cloud visualization of the keywords used during the package search process.}
    \label{fig:wordcloud}
\end{figure}

\noindent \textbf{Keyword-based Package Search.} 
To construct our dataset, we queried the PyPI and NPM platforms for open-source packages associated with LLMSC components. We used a comprehensive set of keywords and their combinatiGenerate a comprehensive list of keywords for searching software packages related to large language models (LLMs) in the NPM and PyPI ecosystems. Include terms covering core functionalities (e.g., inference, fine-tuning, embeddings), ecosystem tools (e.g., vector databases, retrieval-augmented generation), integration workflows (e.g., APIs, orchestration, prompt engineering), and platform-specific contexts (e.g., OpenAI, Hugging Face).ons, including but not limited to ``LLM,'' ``Pre-trained Models,'' ``GPT,'' ``Transformer,'' ``Chatbot,'' ``Agent'', and ``Prompt'', etc. These keywords were selected based on relevant literature and the terminology commonly used in the LLM ecosystem~\cite{wang2024agentsurvey,diaz2024llmops,hadi2025large}. \autoref{fig:wordcloud} highlights the most frequently used keywords in the search process, providing an overview of the terminology that guided our dataset construction.
For PyPI, we developed a crawler that executed keyword searches directly on the PyPI marketplace~\cite{pypi} and retrieved all search results sorted by relevance. The crawler collected metadata fields such as the package name, summary, description, project URLs, and README files. To ensure comprehensive coverage, the crawler iteratively paginated through all search results for each keyword. For NPM, we leveraged the NPM Registry API~\cite{npm} to perform keyword-based searches and retrieved metadata fields including the package name, description, dependencies, and README content. 
Using this process, we collected a total of 54,393 packages, including 43,111 from PyPI and 11,282 from NPM.

\noindent \textbf{LLM-related Packages Filtering.}  
After collecting all packages, we applied a filtering process to identify those specifically related to LLMs. We employed a two-stage heuristic classification method. In the first stage, we leveraged the pre-trained DeepSeek-V3 model~\cite{deepseek} to analyze the README content and metadata of each software package. Packages explicitly mentioning functionalities related to model training, inference, data preprocessing, or orchestration frameworks were labeled as ''LLM-related'' while those whose relevance was unclear were categorized as ''LLM-related-''. 
To address potential classification inaccuracies from the first stage, the second stage involved a standardized manual review process for all packages in the ``uncertain'' category. In this process, the first two authors independently reviewed and labeled the packages based on predefined criteria for LLM relevance, such as functionality related to model training, inference, fine-tuning, data preprocessing, or orchestration frameworks. In cases of disagreement or ambiguity, a third reviewer with advanced expertise was consulted to resolve the conflict and reach a consensus. This systematic approach ensured consistency and reliability in the final classification. During this step, we systematically excluded packages that did not meet the criteria for LLM relevance, such as those pertaining to general-purpose tools or unrelated machine learning functionalities. After applying the filtering process, we curated a final dataset comprising 13,486 LLM-related software packages, with 10,354 originating from PyPI and 3,132 from NPM. To validate the reliability of this pipeline, we randomly sampled 2\% (270) packages for manual verification, achieving an accuracy of 94.8\%. While the validation results suggest a small proportion of inaccuracies, we believe the overall data quality remains sufficiently reliable to support subsequent analysis.

\begin{table}[t]
    \centering
    \caption{Overview of vulnerability processing.}
    \label{tab:cve}
    \fontsize{9}{12}\selectfont
    \begin{tabular}{lrrr}
        \toprule[1.2pt]
        \textbf{Sources}     & \textbf{\texttt{huntr}} & \textbf{GHA} & \textbf{NVD} \\
        \midrule
        Collected CVEs                  & 1,788           & 567            & 4,796          \\
        Deduplication               & -387           & -23            & -2,330         \\
        Package Relevance                   & -1,253          & -520           & -2,458         \\
        \textbf{Final}                       & 148           & 24             & 8           \\
        \bottomrule[1.2pt]
    \end{tabular}
\end{table}

\noindent \textbf{Vulnerability Collection and Processing.}
To comprehensively study vulnerabilities within the LLMSC, we aggregated data from a diverse set of sources, including official databases such as NVD~\cite{nvd} and GitHub Advisory~(GHA) database~\cite{gitadvisory}, and bounty platforms like \texttt{\texttt{huntr}}~\cite{huntr}. Using each package’s repository name as our primary query key, we initially collected a total of 7,151 vulnerability records—4,796 from the NVD CVE list, 567 from the GHA database, and 1,788 from \texttt{huntr}, a bug bounty platform specializing in AI/ML vulnerabilities.
After collecting the raw vulnerability data, we processed it through two filtering steps, as summarized in \autoref{tab:cve}. First, we deduplicated records within and across sources, prioritizing vulnerabilities from \texttt{huntr} for their PoCs and detailed fixes, followed by GHA for its developer reviews, and finally NVD as a supplementary source. This process removed 387 records from \texttt{huntr}, 23 from GHA, and 2,330 from NVD.
Second, we evaluated the relevance of each vulnerability to the LLMSC ecosystem by verifying its association with our curated LLMSC package list. Since vulnerabilities retrieved using package names often included false positives (e.g., unrelated repositories or similarly named software), we cross-checked metadata fields such as affected package names, repository URLs, and ecosystem identifiers (e.g., PyPI or NPM). Only vulnerabilities explicitly tied to packages in the LLMSC list were retained, excluding 1,253 records from \texttt{huntr}, 520 from GHA, and 2,458 from NVD.
After these steps, we curated a dataset of 180 vulnerabilities: 148 from \texttt{huntr}, 24 from GHA, and 8 from NVD.

\begin{algorithm}[t]
	\fontsize{8.5}{9}\selectfont
	\caption{LLMSC Construction Algorithm}
	\label{alg:llmsc}
	\SetKwInput{KwInput}{Input}
	\SetKwInput{KwOutput}{Output}
	\SetKwComment{tcp}{\color{blue}// }{} 
	\SetKwProg{Fn}{Function}{:}{}
        \tcp{\textcolor{blue}{Structured as (package, version) tuples}}
	\KwInput{Local dataset $\mathcal{D}$ , API endpoints $\mathcal{A}$}
	\KwOutput{Constructed dependency graph $\mathcal{G} = (V, E)$}

	$V \gets \emptyset$, $E \gets \emptyset$\; \label{ln:init}
	\ForEach{$(package, version) \in \mathcal{D}$}{ \label{ln:node_creation}
		$V \gets V \cup {CreateNode}(package, version)$\;  \label{ln:add_node}
	}
	\tcp{\textcolor{blue}{Dependency resolution workflow}}
	\Fn{ProcessDependencies($\mathcal{G}$)}{ \label{ln:func_start}
		\ForEach{$n \in V$}{ \label{ln:loop_nodes}
			\eIf{${HasLocalDeps}(n, \mathcal{D})$}{ \label{ln:local_check}
				$E \gets E \cup {LinkExistingDeps}(n, \mathcal{D})$\; \label{ln:local_link}
			}{ \label{ln:remote_start}
				${resp} \gets \mathcal{A}.{fetch}(n.package, n.version)$\; \label{ln:api_call}
				
					$V \gets V \cup {CreateNode}({resp}.package, {resp}.version)$\; \label{ln:remote_node}
					$E \gets E \cup (n \to {resp}.{node})$\; \label{ln:remote_edge}
				
			}
		}
	}

	\tcp{\textcolor{blue}{Package version normalization}}
	\Fn{HandleV2Packages($\mathcal{G}$)}{ \label{ln:unversioned_start}
		\ForEach{$package \in {GetV2}(\mathcal{D})$}{ \label{ln:V2_loop}
			$v_{{latest}} \gets \mathcal{A}.{get\_latest}(package)$\; \label{ln:get_latest}
			$n' \gets {CreateNode}(package, version_{{latest}})$\; \label{ln:create_latest}
			{ProcessDependencies}($n'$)\; \label{ln:recursive_call}
		}
	}
	\tcp{\textcolor{blue}{Orchestration layer}}
	$\mathcal{G} \gets {ProcessDependencies}() \cup {HandleV2Packages}()$\; \label{ln:final_graph}
	\Return $\mathcal{G}$\;
\end{algorithm}

\subsection{LLM Supply Chain Construction}
The LLMSC is conceptualized and structured as a directed graph to capture the intricate dependency relationships within the software ecosystem. In this graph, nodes represent \textit{software packages}, and directed edges represent \textit{package dependency relationships}.
Formally, we define the LLMSC as a directed graph \(G = \langle V, E \rangle\), where \(V\) is the set of nodes, and each node corresponds to a unique package-version pair. \(E\) is the set of directed edges, where each edge \((v_{{from}}, v_{{to}}) \in E\) indicates that the downstream package \(v_{{from}} \in V\) depends on the upstream package \(v_{{to}} \in V\), meaning \(v_{{to}}\) is imported by \(v_{{from}}\).
In addition to this versioned graph, we also consider a \textit{normalized graph}, where nodes represent unique packages (irrespective of version) and only the latest version of each package is included. This normalized graph provides a simplified view of the dependency relationships.
Each package node has its associated metadata, including its vulnerability information. Specifically, we attach a vulnerability count attribute \({vuln}(v)\), which represents the number of known vulnerabilities affecting the package. 

In this context, the \textit{degree} of a node refers to the number of edges connected to it, with distinctions made between \textit{in-degree} and \textit{out-degree}:
\begin{itemize}[leftmargin=15pt]
    \item The \textit{in-degree} of a node \(v\), denoted as \(\deg^-(v)\), is the number of directed edges pointing to \(v\), which represents the number of downstream packages that depend on this package. 
    \item The \textit{out-degree} of a node \(v\), denoted as \(\deg^+(v)\), is the number of directed edges originating from \(v\), which represents the number of upstream packages this package depends on.
\end{itemize}

Using these definitions, we can classify package nodes based on their connectivity in the dependency graph:
\begin{itemize}[leftmargin=15pt]
    \item A \textit{root node} in the package dependency graph is a node with \(\deg^+(v) = 0\) and \(\deg^-(v) > 0\), which represents a package that is depended upon by other packages but does not itself depend on any other packages.
    \item An \textit{isolated node} is a node with both \(\deg^-(v) = 0\) and \(\deg^+(v) = 0\), which represents a package that neither depends on others nor is depended upon.
\end{itemize}

The construction process, summarized in \autoref{alg:llmsc}, begins by parsing package names and their associated version information from a local dataset $\mathcal{D}$. The methodology represents each unique package-version pair as a distinct node in the graph. After initializing the nodes, the process resolves dependencies for each package-version node. For dependencies available in the local dataset, the methodology directly establishes edges to link the package to its dependencies. When dependencies are not found locally, we query the \texttt{Libraries.io} API~\cite{librariesio} to fetch the missing metadata. If the queried dependency is classified as an LLM-relevant package, the process adds a new node to the graph and creates the corresponding edge linking it to the dependent package. For cases where dependency declarations lack specific version information or when dependency data are unavailable locally, we use the latest version of the package as a fallback to mitigate issues caused by incomplete or ambiguous dependency declarations.

\begin{table}[t]
\centering
\caption{Overview of the LLMSC statistics.}
\label{tab:llmsc}
\begin{threeparttable}
\fontsize{8.5}{12}\selectfont
\begin{tabular}{crrrr}
    \toprule[1.2pt]
    \multirow{2}{*}{\textbf{Property}} & \multicolumn{2}{c}{\textbf{Versioned Graph}} & \multicolumn{2}{c}{\textbf{Normalized Graph}} \\
    \cmidrule(lr){2-3} \cmidrule(lr){4-5}
                      & \textbf{PyPI} & \textbf{NPM} & \textbf{PyPI} & \textbf{NPM} \\
    \midrule
    \#\(V_{pkg}\)             & 123,992 & 89,077 & 10,354 & 3,132 \\
    \#\(E_{dep}\)  & 685,660 & 134,421 & 26,492 & 2,212\\
    \#\(depth(G)\)                     & 10 & 8 & 9 & 6 \\
    \#\textit{root node} & 9,247 & 1,277 & 1,713 & 858 \\
    \#\textit{isolated node} & 23,061 & 29,038 & 2,283 & 1,073 \\
    \bottomrule[1.2pt]
\end{tabular}
\end{threeparttable}
\end{table}

The final LLMSC comprises two complementary representations: the \textit{versioned graph} and the \textit{normalized graph}. The versioned graph captures all unique package-version pairs and their version-specific dependencies, resulting in a detailed structure with 123,992 nodes and 685,660 edges for PyPI, and 89,077 nodes and 134,421 edges for NPM. However, this level of granularity introduces a significant number of isolated nodes (51,893 for PyPI and 2,832 for NPM), which are primarily caused by version proliferation and disconnected versions. To address this, the normalized graph consolidates all versions into a single node representing the latest version of each package. This reduces the number of nodes and edges while significantly lowering the number of isolated nodes (to 2,283 for PyPI and 1,073 for NPM), providing a cleaner and more compact representation of the ecosystems. Together, these two representations enable a comprehensive analysis of software dependency structures, balancing fine-grained version-specific details with a simplified view for high-level insights.

%% file: Chapters/3.RQ1.tex
\section{RQ1: Structure}
\label{sec:rq1}
\subsection{Methodology}
To investigate the structure of the LLMSC, we focus on three key aspects: the composition of dependency trees, the topological structures within these trees, and the interconnections between different trees. First, we identify all distinct dependency trees in the LLMSC, treating each as a rooted directed acyclic graph (DAG), where the root node represents the entry package and edges indicate dependency relationships. For each tree, we calculate its depth (longest path from root to leaf) and size (total number of nodes) to analyze their distribution and characterize the overall structure of the LLMSC, reporting properties such as mean, median, and outliers. Second, we examine the topological structures within these trees, classifying subtrees into four main categories: chain (linear sequences), star (central nodes with multiple direct dependents), mesh (densely connected subgraphs), and diamond (nodes with shared intermediate dependencies). Using graph traversal algorithms, we calculate the proportions of these structures and evaluate their impact on dependency resolution and the stability of the ecosystem. Finally, we analyze how different dependency trees interconnect by identifying shared nodes and dependency paths, which serve as bridges between otherwise independent trees. For shared nodes, we calculate their frequency and the average number of trees they belong to, while for shared paths, we measure the fraction of reused paths across the LLMSC.

\subsection{Results}
\textbf{RQ1.1:}
\textit{What are the overall structural characteristics of the LLMSC?} 
To comprehensively analyze the structural characteristics of the LLMSC, we focus on the dependency trees within the graph. Each dependency tree represents a potential dependency chain, modeled as a rooted DAG where root nodes serve as entry points and edges indicate dependency relationships. Through multiple analytical perspectives, we provide a comprehensive examination of the overall structural characteristics of the LLMSC. 

\noindent \textbf{Dependency Tree Size and Distribution.}
As listed in \autoref{tab:dependency_tree_stats}, the LLMSC contains a total of 2,571 dependency trees, ranging from small trees with only a single node to large, complex trees with up to 5,205 nodes. The average size of a dependency tree is 91 nodes, with a median size of 2 nodes. This indicates that while a majority of the trees are small, a few large trees dominate the structure. The size distribution exhibits a heavy-tailed pattern: the majority of trees (72.2\%) contain fewer than 5 nodes, while a small number of large trees contribute disproportionately to the overall node count. Notably, the 100 largest trees alone account for 76.82\% of all nodes. These larger trees often represent highly interconnected components with significant influence on the ecosystem’s behavior.

\noindent \textbf{Dependency Tree Depth and Path Lengths.} 
The depth of a dependency tree represents the longest path from the root to a leaf node, providing insight into the complexity of dependency chains. The deepest tree in the LLMSC has a depth of 9, while the average depth is 1.59. Most trees exhibit shallow depths, with 92.57\% of trees having a depth of 3 or less. However, the existence of deep trees suggests that some packages rely on long dependency chains, which could increase the risk of dependency resolution issues or cascading failures.

\noindent \textbf{Node Degree Distribution and Critical Hubs.}  
The degree distribution of nodes within the dependency trees reveals important structural patterns. Most nodes in the LLMSC have low degree centrality, with 38.22\% of nodes acting as leaf nodes (degree = 1). However, a small number of nodes function as hubs, with degrees as high as 2,267. These hub nodes are critical connectors within the trees, linking multiple subtrees and serving as key points of dependency resolution. Their centrality makes them highly influential, but also potential single points of failure if compromised.


\begin{table}[t]
\centering
\caption{Statistics of dependency trees in the LLMSC.}
\label{tab:dependency_tree_stats}
\fontsize{9}{12}\selectfont
\begin{tabular}{lr}
\toprule[1.2pt]
\textbf{Metric} & \textbf{Value} \\
\midrule
Total number of dependency trees & 2,571 \\
Largest tree size (number of nodes) & 5,205 \\
Smallest tree size (number of nodes) & 2 \\
Average tree size (number of nodes) & 91 \\
Median tree size (number of nodes) & 2 \\
Longest tree depth & 9 \\
Average tree depth & 1.59 \\
Maximum node degree (hub node degree) & 2,265 \\
Percentage of leaf nodes (degree = 1) & 38.22\% \\
\bottomrule[1.2pt]
\end{tabular}
\end{table}


\textbf{RQ1.2:}
\textit{What are the topological substructures within the dependency trees of LLMSC, and how prevalent are they?} 
To investigate the internal composition of dependency trees within the LLMSC, we analyze the topological substructures that shape the ecosystem’s functionality. These substructures provide insight into how dependencies are organized and distributed, revealing patterns of modularity, centralization, and connectivity.

\begin{figure*}[t]
\centering
\includegraphics[width=0.8\textwidth]{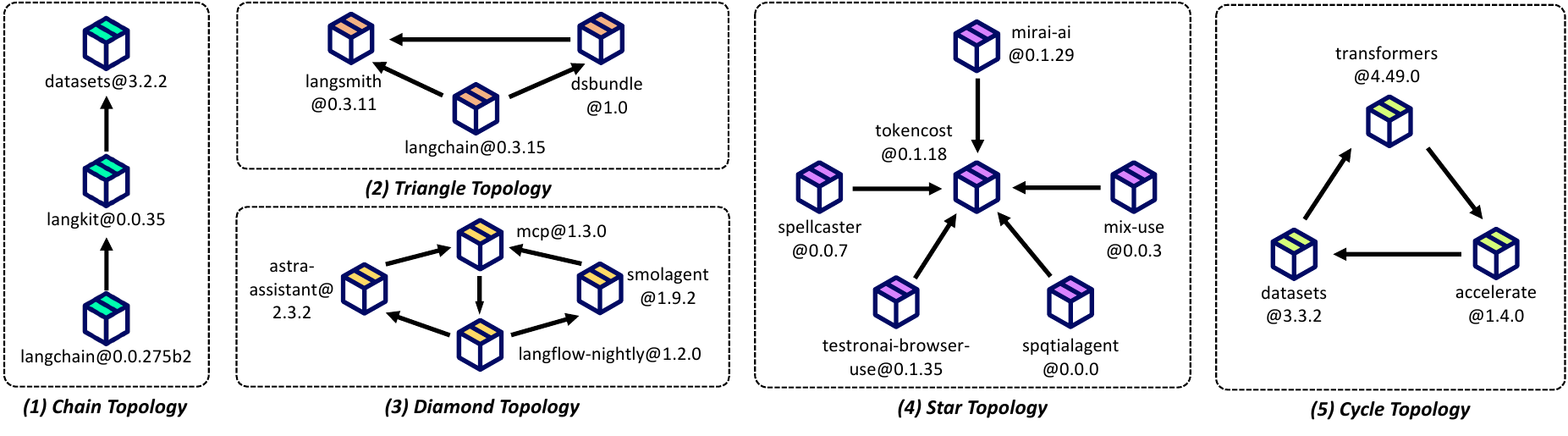}
\caption{Representative examples of topological substructures in dependency trees. An edge pointing from A to B indicates that B depends on A, or equivalently, B is a dependent of A.}
\label{fig:substructure_examples}
\end{figure*}

\noindent \textbf{Formal Definitions of Substructures.}  
To better understand the organization and interaction of packages within the LLMSC dependency trees, we provide formal definitions of several key topological substructures. These substructures capture distinct patterns of connectivity and modularity, offering insights into how dependencies are distributed and reused across the ecosystem. As shown in \autoref{fig:substructure_examples}, these substructures are formally defined as follows:

\begin{definition}[Chain Topology]
A chain is the simplest substructure, representing a strictly linear sequence of dependencies without any branching. Formally, it is defined as a simple path \(P = (v_1, v_2, \ldots, v_n)\), where each node \(v_i\) satisfies:  
\begin{itemize}[leftmargin=15pt]
    \item The first node \(v_1\) and the last node \(v_n\) are leaf nodes, satisfying \(deg(v_1) = deg(v_n) = 1\).  
    \item All intermediate nodes \(v_i\) (\(1 < i < n\)) have exactly one incoming edge and one outgoing edge, satisfying \(deg(v_i) = 2\).  
\end{itemize}
\end{definition}

\begin{definition}[Triangle Topology]
A triangle is a closed loop of three nodes \(v_1, v_2, v_3\), where every pair of nodes is directly connected by an edge. Formally, a triangle is defined as:  
\[
\{(v_1, v_2), (v_1, v_3), (v_2, v_3)\} \subseteq E.
\]  
Triangles indicate tightly coupled dependency clusters, reflecting strong interconnections between components. 
\end{definition}

\begin{definition}[Diamond Topology]
A diamond topology consists of two paths \(P_1 = (v_1, v_2, v_4)\) and \(P_2 = (v_1, v_3, v_4)\), where \(v_2 \neq v_3\). Formally, a diamond is defined as:  
\[
\{(v_1, v_2), (v_2, v_4), (v_1, v_3), (v_3, v_4)\} \subseteq E.
\]  
Diamonds are prevalent in shared libraries, where intermediate dependencies are reused across multiple packages. 
\end{definition}

\begin{definition}[Star Topology]
A star topology is defined as a central node \(v_c\) connected to a set of leaf nodes \(v_1, v_2, \ldots, v_k\), where:  
\[
deg(v_c) = k, \quad deg(v_i) = 1 \ \text{for all} \ i = 1, \ldots, k.
\]  
Stars are typical of centralized libraries that provide core functionality to multiple dependent components. 
\end{definition}

\begin{definition}[Cycle Topology]
A cycle represents a circular dependency, where a closed path \(C = (v_1, v_2, \ldots, v_k, v_1)\) exists, and every node \(v_i\) is connected to exactly two others. Formally, a cycle is defined as:  
\[
\{(v_i, v_{i+1}) \ | \ i = 1, \ldots, k-1\} \cup \{(v_k, v_1)\} \subseteq E.
\]  
Cycles often signify complex interdependencies, which can make dependency resolution challenging. 
\end{definition}

\begin{table}[t]
\centering
\caption{Statistics of sub-topological structures in LLMSC dependency trees.}
\label{tab:substructure_stats}
\fontsize{9}{12}\selectfont
\begin{tabular}{lrc}
\toprule[1.2pt]
\textbf{Typology} & \textbf{Count} & \textbf{Example} \\
\midrule
Chain Typology & 491 & datasets@3.2.2 \\
Triangle Typology & 14,992 & langsmith@0.3.11 \\
Diamond Typology & 35,667 & mcp@1.3.0 \\
Star Typology & 1,362 & tokencost@0.1.18 \\
Cycle Typology & 619 & transformers@4.49.0 \\
\bottomrule[1.2pt]
\end{tabular}
\end{table}

\noindent \textbf{Prevalence of Substructures.}  
Dependency trees in the LLMSC are composed of these primary sub-topological structures, with their prevalence summarized in \autoref{tab:substructure_stats}. Among these, diamonds are the most prevalent, with 35,667 instances, primarily representing shared libraries where intermediate dependencies are extensively reused. Triangles are the second most common structure, accounting for 14,992 instances, reflecting tightly coupled dependency clusters.  
Star structures, with 1,362 instances, indicate highly centralized topologies where core packages directly support multiple dependent components. Chains, totaling 491 instances, typically occur in smaller or simpler dependency trees characterized by sequential, linear relationships. Notably, cycle structures—though often considered atypical—are present in 619 cases, revealing non-negligible circular dependencies that complicate version resolution and heighten the risk of cascading failures.  


\textbf{RQ1.3:}
\textit{How are the various dependency trees interconnected within the LLMSC?}  
The interconnectedness of dependency trees within the LLMSC arises from shared dependency nodes, common dependency paths, and varying degrees of tree coupling. These interconnections are essential for understanding the supply chain’s structural characteristics, as they highlight both the modularity and the potential propagation pathways for vulnerabilities across projects. Here, we analyze these factors in detail and summarize the key findings.

\begin{table}[t]
\centering
\caption{Summary of interconnection metrics.}
\label{tab:interconnection_metrics}
\fontsize{9}{12}\selectfont
\begin{tabular}{lr}
\toprule[1.2pt]
\textbf{Metric} & \textbf{Value} \\
\midrule
\rowcolor[gray]{0.9}
Total number of shared nodes & 6,496 \\
Nodes shared by exactly 2 trees & 975 \\
Nodes shared by 3–10 trees & 3,216 \\
Nodes shared by more than 10 trees & 2,305 \\
\hline
\rowcolor[gray]{0.9}
Total number of shared dependency paths &10,489 \\
Paths shared by 2 trees & 44.68\% \\
Paths shared by 3–5 trees & 35.11\% \\
\hline
Tree pairs with at least one shared node & 62,385 \\
Tree pairs sharing fewer than 10 nodes & 47,800 \\
Tree pairs sharing more than 50 nodes & 5,013 \\
\bottomrule[1.2pt]
\end{tabular}
\end{table}

\begin{figure}[t]
    \centering
    \includegraphics[width=0.85\linewidth]{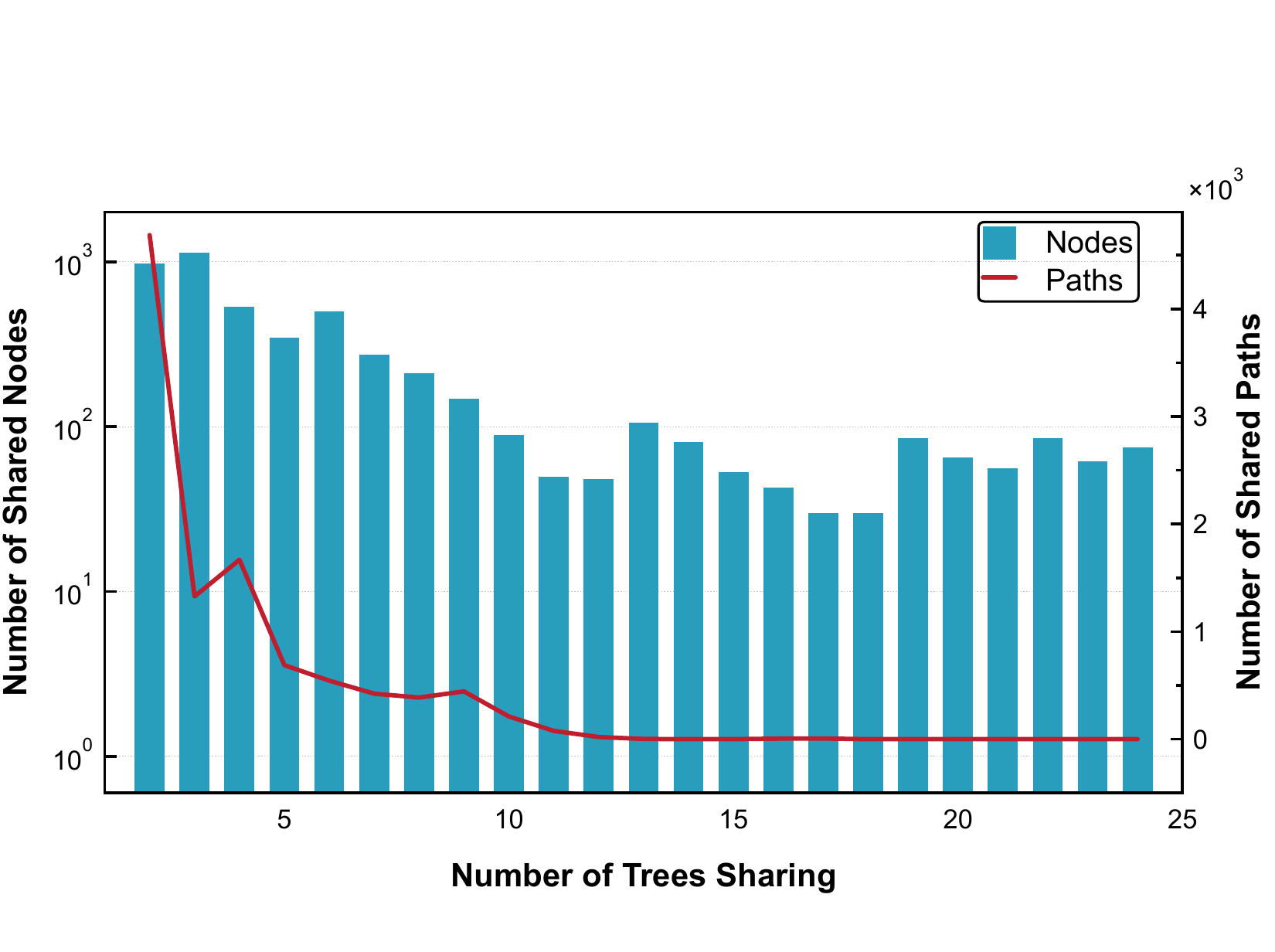}
    \caption{Distribution of shared dependency nodes and paths.}
    \label{figure:shared_distribution}
\end{figure}

\noindent \textbf{Shared Dependency Nodes.}  
Shared nodes are components that appear in multiple dependency trees, serving as bridges that connect different projects. From our analysis, a total of 6,496 nodes are shared across at least two dependency trees, with the majority shared by fewer than 10 trees (accounting for 62.8\% of all shared nodes). As shown in \autoref{figure:shared_distribution}, the distribution of shared nodes follows a long-tailed pattern. Although the majority are shared across fewer than 10 trees , with 1,132 appearing in exactly three trees and 531 in four, the cumulative count of nodes shared across more than 10 trees reaches 2,305, indicating the existence of widely adopted, ecosystem-level dependencies. These highly shared nodes often represent critical libraries or frameworks, such as the chatgpt-academic@3.60.3 (shared across 50 trees) and llm-eval-toolkit@1.0.4 (shared across 43 trees). These components form the backbone of the ecosystem by providing essential, reusable functionality. However, their centrality also makes them potential single points of failure. 

\noindent \textbf{Shared Dependency Paths.}  
Shared dependency paths, defined as sequences of dependencies that are reused across multiple trees, reflect common functional stacks or workflows within the LLMSC. Our analysis identified 10,489 shared paths of length greater than two (i.e., containing at least three nodes). Most of these paths (44.68\%) are shared by exactly two trees, while a smaller percentage are shared among three to five trees (35.11\%) . Only a handful of paths (2119) are shared by more than five trees. These shared paths often correspond to specific frameworks or workflows that are widely adopted across projects. For example, the sequence ``Interpret → Error‑Analysis → Model‑Assessment'' is a recurring subpath in Responsible AI toolchains, while the chain ``GraphQL Transformer → GraphQL API Construct → Data Construct'' is frequently reused in projects on the AWS Amplify platform. \autoref{figure:shared_distribution} illustrates the distribution of shared paths across trees, highlighting the concentration of path reuse in a small subset of workflows. Such shared paths represent reusable templates but also create interdependencies that could propagate issues if any part of the path fails or becomes incompatible.

\begin{tcolorbox}
    \textbf{Answer to RQ1.} The LLMSC consists of 2,571 dependency trees, 72.2\% of which have fewer than 5 nodes, while a few large trees and high-degree hubs dominate. Internally, diamond (67.13\%) and triangle (28.22\%) topologies are the main substructures. Moreover, 6,496 shared nodes and 10,489 shared paths indicate interconnections and potential risk across trees.
\end{tcolorbox}




%% file: Chapters/4.RQ2.tex
\section{RQ2: Domain Distribution}
\label{sec:rq2}
\subsection{Methodology}
To investigate the domain composition of the LLMSC and its evolution, we focus on three key aspects: the static distribution of functional domains, the temporal evolution of these domains, and the hierarchical transitions of domains within dependency trees. First, we assign each package to a functional domain, such as ``Data Pipelines,'' ``Training Frameworks,'' and ``Inference Tools,'' based on its metadata and README content using a combination of keyword matching and the pre-trained classification model DeepSeek-V3~\cite{deepseek}. Then, we analyze hierarchical domain transitions within dependency trees, treating each as a rooted directed acyclic graph (DAG) and tracing the flow of domains from root to leaf nodes. These analyses provide a comprehensive understanding of the LLMSC’s domain-level composition, its evolution over time, and the structural relationships between functional domains.

\subsection{Results}
\textbf{RQ2.1:} \textit{What is the domain distribution of the LLMSC, and how has it evolved over time?}
To understand the domain composition of the LLMSC and its temporal evolution, we analyzed the functional domains of these packages and tracked the growth trends across these domains over the past decade. The key findings are as follows:

\noindent \textbf{Domain Classification.} To analyze the domain distribution of the LLMSC, we first classified the retrieved packages using a heuristic domain inspired by prior research~\cite{wang2025sok}. This framework defines multiple stages in the domain and technical stack, including data preprocessing, model training, deployment, and post-deployment monitoring, each of which is supported by specialized tools and frameworks. Based on this lifecycle, we categorized all packages into 14 functional domains, such as ``Data Pipeline,'' ``Training and Fine-tuning Frameworks,'' ``LLM Inference,'' and ``Orchestration Frameworks.'' To assign packages to these domains, we employed DeepSeek-V3~\cite{deepseek}, a pre-trained classification model that processes the README content and metadata of each package. The model automatically classified packages into the predefined categories based on their described functionalities, ensuring that each package was assigned to the most relevant domain.

\begin{table}[t]
    \centering
    \caption{Domain distribution of LLMSC packages.}
    \label{tab:domain_distribution}
    \fontsize{9}{12}\selectfont
    \begin{tabular}{lrr}
        \toprule[1.2pt]
        \textbf{Domain} & \textbf{Rate} & \textbf{Count}\\
        \midrule
        Data Index                & 4.66\% & 628 \\
        Data Pipeline             & 4.78\% & 645 \\
        Vector Database           & 4.99\% & 673 \\
        Model Merge               & 0.07\% & 9 \\
        Model Quantization        & 0.47\% & 63 \\
        Training/Fine-tuning Frameworks & 7.85\%   & 1,058 \\
        LLM Gateway               & 23.47\% & 3,165 \\
        LLM Cache                 & 0.90\% & 122 \\
        LLM Inference             & 10.65\%  & 1,436 \\
        Logging\&LLMOps           & 7.09\% & 956 \\
        Orchestration Frameworks  & 11.37\% & 1,533 \\
        RAG Frameworks            & 5.77\%  & 778 \\
        Plugins/External Tools    & 4.78\% & 644\\
        Apps/Front-end Frameworks & 13.17\% & 1,776 \\
        \midrule
        \textbf{Total}            & \textbf{/}   & \textbf{13,486} \\
        \bottomrule[1.2pt] 
    \end{tabular}
\end{table}

\noindent \textbf{Domain Distribution.} The static distribution, as shown in \autoref{tab:domain_distribution}, indicates that the LLMSC is primarily driven by a few key domains. ``LLM Gateway'' account for the largest proportion, comprising \( 23.47\% \) (3,165 packages) of the ecosystem,  underscoring its central role in mediating interactions between applications and LLM backends. ``Orchestration Frameworks'' constitute the second-largest domain at \( 11.37\% \) (1,533 packages), reflecting growing demand for systems that coordinate complex LLM workflows. ``LLM Inference'' follows closely at (\( 10.65\% \), 1,436 packages), highlighting continued emphasis on efficient deployment of pre-trained models.Core development infrastructure is further supported by ``Training and Fine-tuning Frameworks'' (\( 7.85\% \), 1,058 packages), and ``Logging LLMOps'' (\( 7.09\% \), 956 packages),  which address model adaptation and operational observability, respectively. Notably, application-layer domains such as ``Apps/Front-end Frameworks'' represent a substantial share (\( 13.17\% \), 1,776 packages),indicating active efforts to build user-facing LLM-powered interfaces. Specialized technical domains—including ``RAG Frameworks'' (\( 5.77\% \), 778 packages), ``Vector Database'' (\( 4.99\% \), 673 packages), and ``Data Pipeline'' (\( 4.78\% \), 645 packages), play critical roles in retrieval-augmented generation, semantic storage, and data preprocessing. In contrast, low-level optimization techniques like ``Model Quantization'' (\( 0.47\% \), 63 packages) and ``Model Merge'' (\( 0.07\% \), 9 packages), remain niche, suggesting limited tooling maturity in these areas. Emerging paradigms such as “LLM Agent” and “LLM DSL/Programming” are not represented in this snapshot, implying they have yet to crystallize into established open-source categories.

\begin{figure}[t]
    \centering
    \includegraphics[width=0.9\linewidth]{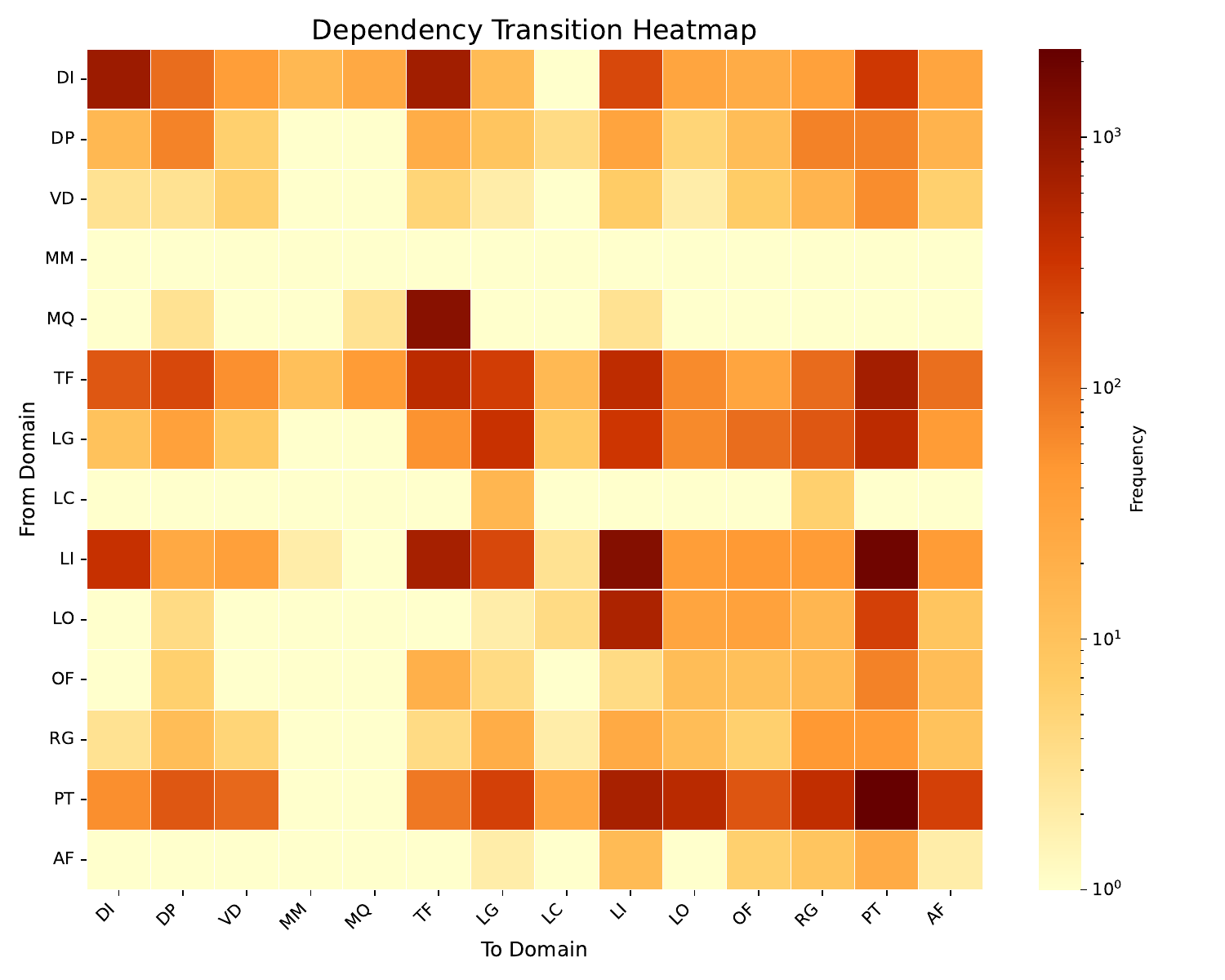}
    \caption{Domain transition heatmap. The heatmap visualizes the pairwise dependencies between domains in the LLMSC, with darker colors indicating higher transition frequencies. Domains are represented by their initials.}
    \label{fig:domain_transition_heatmap}
\end{figure}

\textbf{RQ2.2:} \textit{How do domains transition along root‑to‑leaf paths?}
To investigate how domains transition along root-to-leaf paths in the LLMSC dependency graph, we analyzed 824 unique paths and visualized the pairwise dependencies between domains using a domain transition heatmap, as shown in \autoref{fig:domain_transition_heatmap}. The heatmap highlights the frequency of dependencies between domain pairs, where a transition \textit{From A to B} indicates that domain B depends on domain A. This provides insights into how domains interconnect to support workflows within the ecosystem. 
The analysis reveals that the most common dependencies occur between \textit{LLM Inference} and \textit{Plugins / External Tools}, with \textit{Plugins / External Tools} frequently depending on \textit{LLM Inference}. This reflects the reliance of plugin tools on inference frameworks for core functionalities such as executing pre-trained models or integrating inference pipelines. Another highly recurrent dependency is \textit{Plugins / External Tools $\rightarrow$ Plugins / External Tools}, which underscores the ecosystem's modular nature, where plugins often depend on other plugins to extend their capabilities. 

\begin{tcolorbox}
    \textbf{Answer to RQ2.} 
    The LLMSC is dominated by \textit{LLM Gateway} (\(23.47\%\)), which acts as a central orchestration layer mediating interactions between applications and LLM backends. Our analysis reveals that common workflow paths frequently traverse \textit{LLM Inference}, \textit{Orchestration Frameworks}, and \textit{Training Frameworks}, underscoring their pivotal roles in the overall workflow.
\end{tcolorbox}

%% file: Chapters/5.RQ3.tex
\section{RQ3: Security Risks}
\label{sec:rq3}
\subsection{Methodology}
Building on the structural analysis of the LLMSC in \autoref{sec:rq1}, RQ3 investigates how vulnerabilities propagate within the supply chain and identifies structural factors that contribute to security risks. To achieve this, we analyzed the characteristics of nodes to determine which nodes or dependency structures are more prone to security issues. Specifically, we focused on nodes with extensive upstream dependencies, which are more likely to experience cascading failures when vulnerabilities occur, and nodes with a large number of downstream dependencies, where vulnerabilities can propagate widely. Additionally, we examined critical nodes that bridge multiple dependency clusters, as they can amplify risks by acting as single points of failure. Furthermore, we conducted a fine-grained study of vulnerability propagation. By constructing detailed dependency trees for specific vulnerable package versions, we evaluated the downstream impact of vulnerabilities originating in upstream packages. This allowed us to identify which downstream packages are directly or transitively affected and to quantify the extent of the impact. 

\subsection{Results}

\begin{figure}[t]
    \centering
    \begin{subfigure}[b]{0.48\linewidth}
        \centering
        \includegraphics[width=\linewidth]{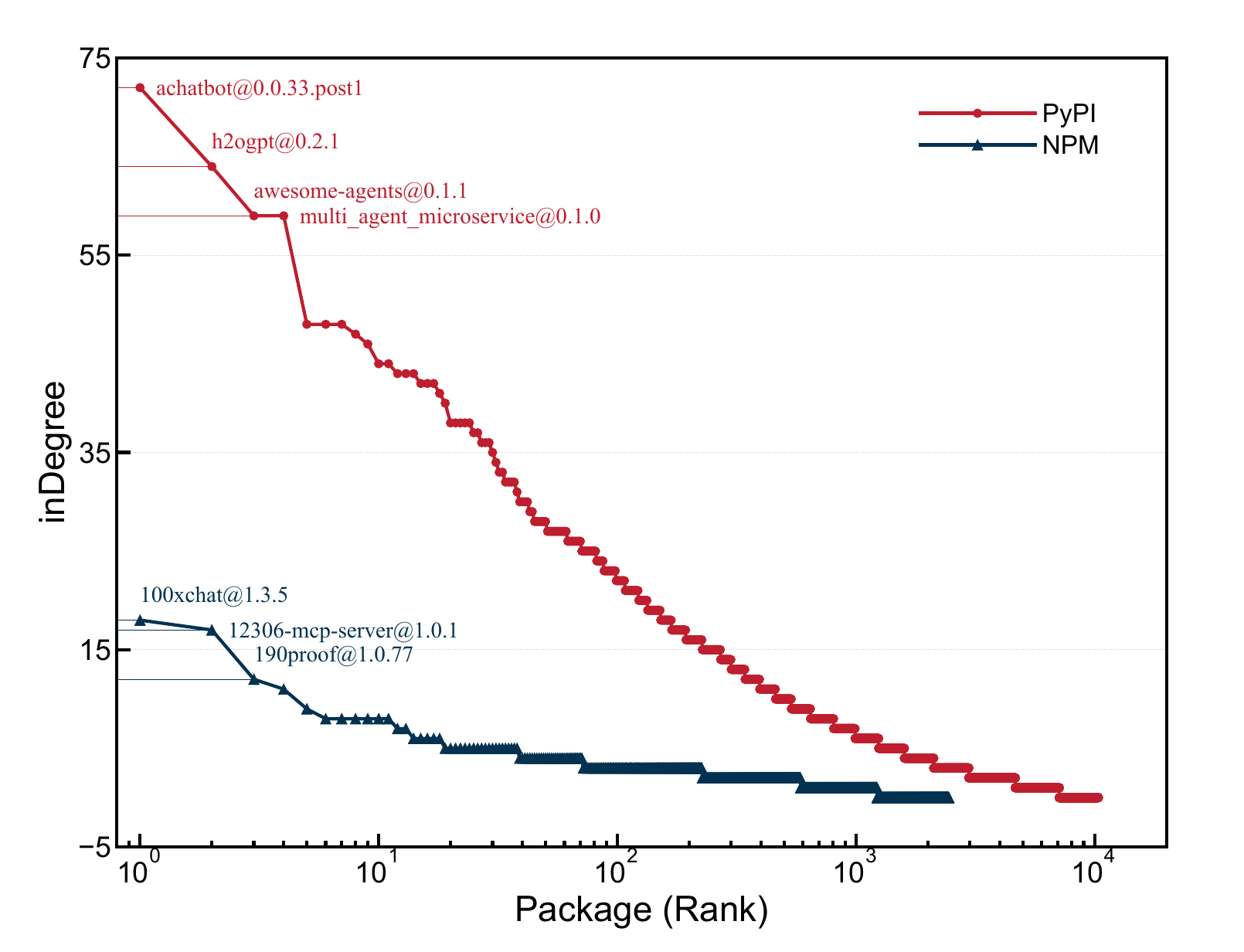}
        \caption{In-degree Distribution}
        \label{fig:indegree}
    \end{subfigure}
    \hfill 
    \begin{subfigure}[b]{0.48\linewidth}
        \centering
        \includegraphics[width=\linewidth]{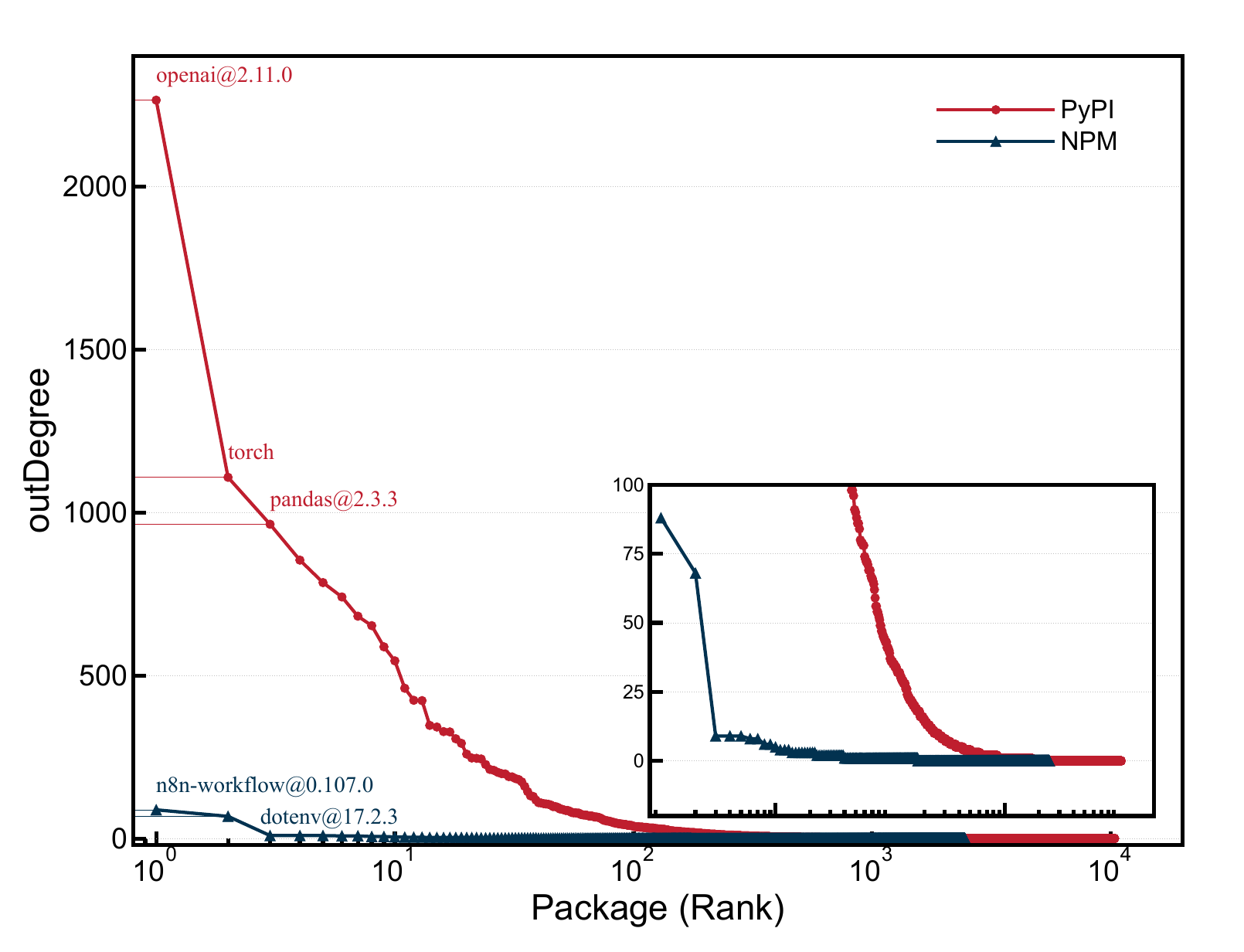}
        \caption{Out-degree Distribution}
        \label{fig:outdegree}
    \end{subfigure}
    \caption{Comparison of in-degree and out-degree distributions in PyPI and NPM.}
    \label{fig:degree_comparison}
\end{figure}

\textbf{RQ3.1:} \textit{Which nodes in the LLMSC qualify as high‑risk, and what security vulnerabilities do they present?}
In the LLMSC, critical nodes are packages that occupy influential positions within the network topology and carry a significant number of dependencies. Identifying these nodes is essential for understanding supply chain security risks and designing effective mitigation strategies. Through a multidimensional analysis using network metrics such as \textit{In-degree}, \textit{Out-degree}, and \textit{Composite Centrality}, we systematically identified high-risk nodes and assessed their security implications.

\noindent \textbf{Nodes with the Most External Dependency.} 
In-degree represents the number of external components a package depends on at runtime. A high in-degree indicates that a package relies heavily on third-party libraries, each of which could introduce potential risks or vulnerabilities.  
By analyzing package dependencies (see \autoref{fig:indegree}), we observe that approximately 85.1\% of packages depend on zero or one component—exhibiting minimal exposure—while only 2.1\% qualify as ``highly exposed'' (in-degree $\ge$ 5). Notably, in the PyPI ecosystem, \texttt{langflow-nightly@1.2.0} leads with 19 dependencies, followed by \texttt{h2ogpt@0.2.1} at roughly 17. The top five packages average an in-degree of 14.7, 18.3 times the network average of 0.8011, implying that each additional dependency compounds potential attack surfaces, and that highly interconnected packages often struggle to keep pace with timely patching.

\noindent \textbf{Nodes with the Most Dependents.}  
Out-degree represents the number of downstream consumers that directly depend on a package. A high out-degree indicates that a package is widely used as a dependency, amplifying its potential impact radius and making it a critical hub in the ecosystem.  
As shown in \autoref{fig:outdegree}, the majority of packages in the LLMSC have low out-degree, with approximately 99.3\% of packages having fewer than 5 dependents. However, a small subset of high out-degree hubs stands out, accounting for a disproportionate share of downstream dependencies.  
For instance, \texttt{openai@4.96.0} has the highest out-degree with 1,891 direct dependents, followed closely by \texttt{pandas@2.2.3} (1,864 dependents) and \texttt{transformers@4.49.0} (1,043 dependents). Collectively, the top five high out-degree packages account for 37.3\% of all downstream chains, demonstrating their critical role in the LLMSC ecosystem. These packages also exhibit substantial indirect dependencies, with \texttt{openai@4.96.0} impacting approximately 2,485 downstream nodes across multiple levels of the dependency tree. The average maximum chain depth for these packages is only 1.47, suggesting that vulnerabilities in these hubs may propagate extensively through the network.

\begin{figure}[t]
    \centering
    \begin{subfigure}[t]{0.48\linewidth}
        \centering
        \includegraphics[width=\linewidth]{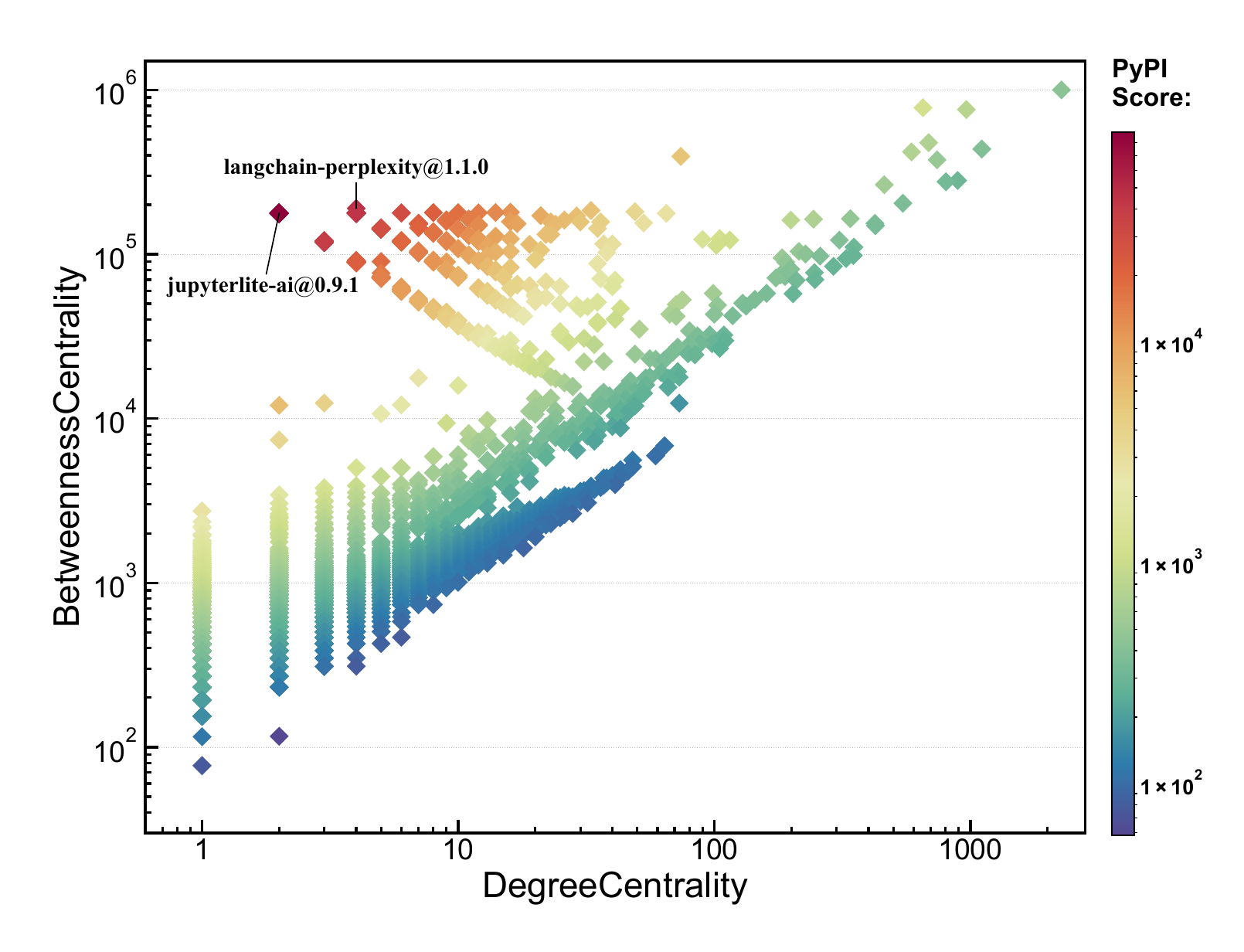}
        \caption{Composite Centrality of PyPI}
        \label{fig:pypi_centrality}
    \end{subfigure}
    \hfill 
    \begin{subfigure}[t]{0.48\linewidth}
        \centering
        \includegraphics[width=\linewidth]{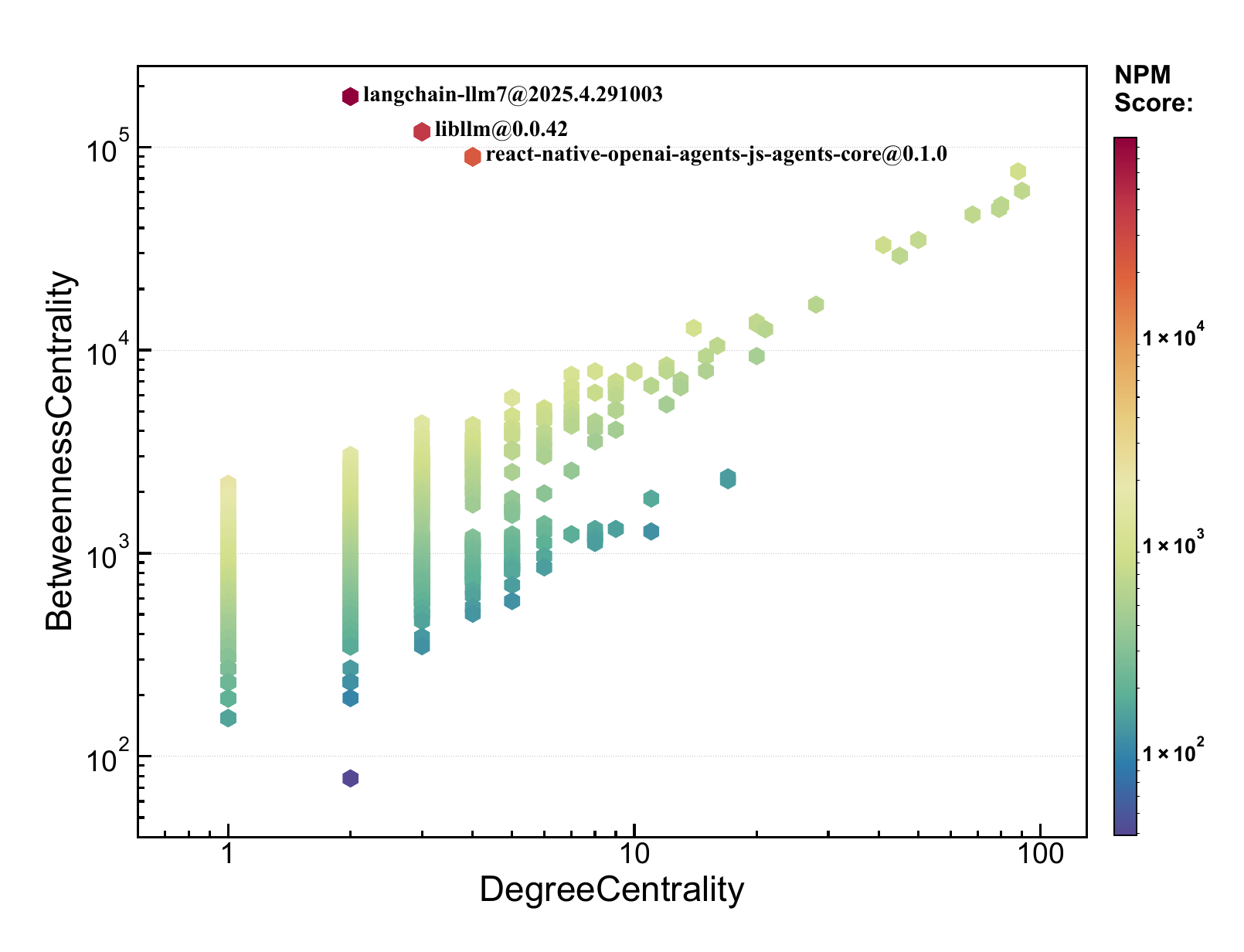}
        \caption{Composite Centrality of NPM}
        \label{fig:npm_centrality}
    \end{subfigure}
    \caption{Comparison of Composite Centrality in PyPI and NPM ecosystems.}
    \label{figure:centrality_comparison}
\end{figure}
\noindent \textbf{Nodes Serving as the Bridges.}
Bridge nodes link otherwise separate dependency clusters, serving as critical conduits for both functionality and potential risk. These Bridge nodes play a pivotal role in the ecosystem by connecting different clusters of dependencies. However, their critical position also makes them vulnerable to cascading risks in the event of a compromise.
To quantify the importance of bridge nodes, we use a metric called \emph{Composite Centrality}, defined as the ratio of \emph{Betweenness Centrality} (BC) to \emph{Degree Centrality} (DC):  
\[
  \mathrm{Composite\ Centrality} = \frac{\mathrm{Betweenness\ Centrality}}{\mathrm{Degree\ Centrality}},
\]
where a high BC/DC ratio indicates that a node lies on many shortest paths between other nodes (high BC) despite having relatively few direct connections (low DC).  
\autoref{figure:centrality_comparison} illustrates the distribution of Composite Centrality in PyPI and NPM ecosystems. These figures demonstrate that NPM exhibits a smaller number of highly central nodes compared to PyPI, indicating differences in ecosystem structure.
For example, \texttt{language-model-toolkit@0.1} in PyPI exhibits a composite centrality score of 42,945, ranking it among the top bridge nodes in the ecosystem, with a betweenness centrality of 85,891 and a degree centrality of 2. This node acts as a critical bridge, connecting otherwise separate dependency trees associated with \texttt{openai@4.96.0}, \texttt{pandas@2.2.3}, \texttt{tiktoken@0.9.0}, and \texttt{onnxruntime-tools@1.7.0}, thereby amplifying potential cascading risks in the event of a compromise.
Nodes with a composite centrality score above the network average are classified as bridge nodes, as they play a disproportionately important role in connecting otherwise separate dependency clusters.
Overall, our analysis identifies 454 bridge nodes in the LLMSC, accounting for approximately 2.9\% of all nodes. 
The average Composite Centrality for bridge nodes is 783, significantly higher than the network average of 81.
These bridge nodes are distributed across key domains such as LLM Inference (14.3\%), Data Pipeline (11.9\%), and Plugins / External Tools (11.7\%).

\textbf{RQ3.2:} \textit{What known vulnerabilities currently exist within the LLMSC? How do vulnerabilities propagate through the dependency trees?}
To address RQ3.2, we conducted an in-depth analysis of known vulnerabilities within the LLMSC and their propagation patterns through dependency trees. Using a dataset of reported CVEs in the LLMSC ecosystem, we analyzed the direct and transitive impact of these vulnerabilities across different layers of dependency trees. We quantified the propagation of vulnerabilities by examining the number of affected nodes at each layer of the dependency tree, from the root package (layer 0) to its transitive dependents (layer \(n\)). 

\begin{table}[t]
\centering
\caption{Average number of nodes affected at each layer (Avg: Average, Std: Standard Deviation, Med: Median).}
\label{tab:average_impact}
\fontsize{9}{12}\selectfont
\begin{tabular}{llrrrr}
    \toprule[1.2pt]
    \textbf{Impact Type} & \textbf{Layer} & \textbf{Avg} & \textbf{Std} & \textbf{Med} & \textbf{Max} \\
    \midrule
    Direct Impact & Layer 1 & 8.3 & 16.0 & 3 & 59 \\
    \cmidrule(lr){1-6}
    \multirow{4}{*}{Indirect Impact} 
    & Layer 2 & 142.1 & 271.1 & 5 & 1,043 \\
    & Layer 3 & 237.8 & 245.9 & 31 & 580 \\
    & Layer 4 & 54.1 & 34.4 & 82 & 82 \\
    & Layer 5 & 13 & 5.7 & 17 & 17 \\
    \bottomrule[1.2pt]
\end{tabular}
\end{table}

\noindent \textbf{Known Vulnerabilities and Their Propagation.}  
We analyzed 180 known CVEs across the LLMSC ecosystem to understand how vulnerabilities propagate through dependency trees. These vulnerabilities primarily originate in root packages and propagate to downstream nodes, demonstrating the cascading risks posed by security issues in upstream dependencies. \autoref{tab:average_impact} summarizes the average number of nodes affected at each dependency layer. On average, vulnerabilities directly impact \(8.3\) nodes at the root layer (layer 1), with a standard deviation of \(16.0\). The propagation impact increases significantly at layer 2, where an average of \(142.1\) nodes are affected, with a maximum impact of \(1,043\) nodes. At layer 3, the average number of affected nodes peaks at \(237.8\), before diminishing at deeper layers, with \(54.1\) and \(13\) nodes affected on average at layers 4 and 5, respectively. This reflects the cascading nature of vulnerabilities in high-impact packages, with the most significant effects concentrated in the early layers of the dependency tree.
One notable example is \texttt{CVE-2023-6730}, a critical deserialization vulnerability in the widely-used \texttt{transformers} package (versions prior to 4.36.0). This vulnerability highlights the potential for extensive propagation due to its central role in the ecosystem. It directly impacts \(16\) root-level packages and propagates to a total of \(1,336\) downstream nodes across multiple layers. Such vulnerabilities in critical packages amplify the risks of cascading failures, particularly when they occur in high-impact hubs. 

\begin{figure}[t]
    \centering
    \includegraphics[width=0.8\linewidth]{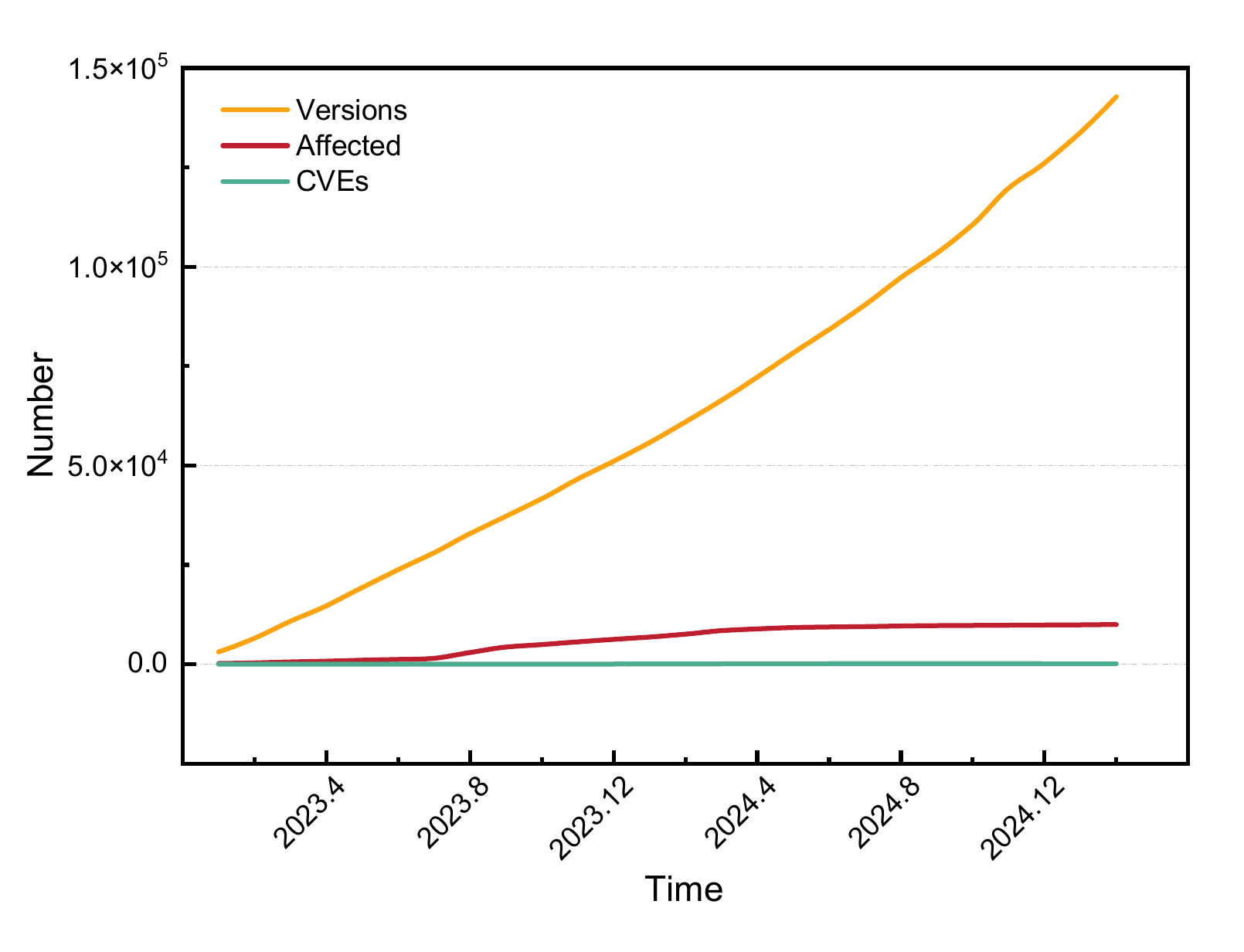}
    \caption{Temporal trends of vulnerabilities in the LLMSC, showing the number of CVEs, affected nodes, and versioned nodes over time (2023.1 to 2025.2).}
    \label{fig:time_series_cves}
\end{figure}

\noindent \textbf{Temporal Analysis of CVEs.}  
From \autoref{fig:time_series_cves}, we observe a steady increase in CVE numbers, affected nodes, and affected versioned nodes over time, highlighting the growing complexity of the LLMSC. The number of CVEs rose from 0 in early 2023 to 159 by 2025.2, with a sharp jump between 2024.4 and 2025.1. Similarly, the number of affected nodes grew rapidly, reaching \(9,885\) by 2025.1, reflecting the cascading impact of vulnerabilities in critical packages. The number of affected versioned nodes also showed consistent growth, surpassing \(100,000\) in 2024.9 and reaching \(142,747\) by 2025.2. These trends emphasize the increasing risks posed by vulnerabilities as the ecosystem expands and diversifies.

\begin{tcolorbox}
\textbf{Answer to RQ3.}  
The LLMSC exhibits significant security risks due to the cascading nature of vulnerabilities. Our analysis of 180 CVEs reveals that vulnerabilities primarily propagate during the first three transmissions in the dependency tree, with an average of \(8.3\), \(142.1\), and \(237.8\) nodes affected, respectively.
\end{tcolorbox}

%% file: Chapters/6.discussion.tex
\section{Discussion}
\label{sec:discussion}
\noindent \textbf{Threats to Validity.}  
One potential threat to the validity of our study stems from the classification of software packages into lifecycle-related categories. We used DeepSeek-V3 to determine the most relevant lifecycle phase for each package, but some packages span multiple stages, and our approach simplified this by assigning them to a single phase. Furthermore, the LLM used for classification may occasionally produce hallucinations, leading to errors, though manual reviews of a subset of classifications showed high consistency. Another limitation lies in our analysis of vulnerability propagation. As noted in prior studies~\cite{wu2023mavensc,wu2024vulpropa,mir2023propagation}, the presence of a vulnerable package in a dependency tree does not guarantee an impact on downstream projects. Vulnerabilities propagate at the function level, and the risk materializes only if the vulnerable code is explicitly invoked downstream. Despite these limitations, our study offers valuable insights into the structure, composition, and vulnerabilities of the LLMSC.

\noindent \textbf{Comparison to Traditional Software Supply Chain.}
The LLMSC differs significantly from traditional software supply chains in several key ways, offering unique insights into its structure and risks. Unlike traditional ecosystems, which often center around a small number of stable core frameworks like TensorFlow~\cite{tensorflow} or PyTorch~\cite{pytorch}, the LLMSC is highly heterogeneous, comprising a diverse range of components tailored to specific stages of LLM development, such as dataset curation, prompt engineering, fine-tuning, and inference. Additionally, the LLMSC is highly dynamic, with new tools and frameworks emerging at an unprecedented pace, driving rapid domain evolution and migration. This fast-paced evolution heightens the complexity of managing dependencies and security risks. Furthermore, the LLMSC’s ``locally dense, globally sparse'' topology amplifies risk concentration in a small set of core libraries that support numerous downstream projects, making them critical points of failure.

%% file: Chapters/7.literature.tex
\section{Related Works}
\label{sec:relatedworks}

\noindent \textbf{Software Supply Chain.}
The study of software supply chains has gained significant attention due to their critical role in modern software development. Research has investigated the structures and dynamics of dependency networks in ecosystems such as PyPI, NPM, Maven, and others. Liu \emph{et al.}~\cite{liu2022npmsc} conducted a large-scale study of the NPM ecosystem, highlighting the propagation of vulnerabilities through transitive dependencies and proposing techniques to resolve dependency trees more accurately. Similarly, Mir \emph{et al.}~\cite{mir2023propagation} analyzed the impact of transitivity and granularity on vulnerability propagation within the Maven ecosystem, emphasizing the importance of fine-grained analysis to identify actual vulnerability paths. Zhang \emph{et al.}~\cite{zhang2023maven} further addressed the persistence of vulnerabilities in Maven by proposing an automated restoration mechanism for secure dependency ranges to mitigate ecosystem-wide risks. Beyond these ecosystems, Tang \emph{et al.}~\cite{tang2023csc} explored dependency patterns in the C/C++ ecosystem, revealing unique challenges due to the lack of a unified package manager. 
Emerging domains such as Web3 also present new challenges to supply chain security. Ma \emph{et al.}~\cite{ma2024web3sc} conducted an in-depth analysis of the Web3 package supply chain, using knowledge graphs to uncover evolving dependencies and vulnerabilities. These studies collectively highlight the vulnerabilities, dependency complexities, and ecosystem-specific challenges in software supply chains.  
In contrast, our work focuses on the unique characteristics of the supply chains associated with LLMs, which have been less explored in prior research.

\noindent\textbf{Deep Learning and Pre-Trained Model Supply Chain.}
The rise of deep learning (DL) frameworks and pre-trained models (PTMs) has introduced unique complexities to software supply chains. Tan \emph{et al.}~\cite{tan2022dlsc} analyzed the supply chains of TensorFlow and PyTorch, unveiling their short, sparse hierarchical structures and revealing the rapid growth of downstream projects dependent on these frameworks. Gao \emph{et al.}~\cite{gao2024dlsc} extended this work by characterizing DL package supply chains in PyPI, identifying domain specializations and dependency clusters critical to the ecosystem's maintenance and evolution. 
Focusing on PTMs, Jiang \emph{et al.}~\cite{jiang2024peatmoss} introduced the PeaTMOSS dataset, which captures metadata for PTMs and their downstream applications, revealing inconsistencies in licensing and documentation practices across the PTM ecosystem. Stalnaker \emph{et al.}~\cite{stalnaker2025mlsc} emphasized similar challenges in the Hugging Face ecosystem, highlighting the need for better documentation, licensing mechanisms, and compliance measures. These studies collectively underscore the importance of addressing dependency complexities, documentation gaps, and emerging risks in DL and PTM supply chains.  
Unlike prior works, our study focuses specifically on the structural, domain-specific, and security characteristics of the LLM toolchain, which remains underexplored.

\noindent\textbf{Supply Chain Security Risks.}
Security vulnerabilities in software supply chains have been a persistent concern, especially in the context of dependency management. Alfadel \emph{et al.}~\cite{alfadel2023pypi} analyzed vulnerabilities in Python packages, observing that vulnerabilities often remain unfixed for extended periods, leaving downstream projects exposed. Huang \emph{et al.}~\cite{huang2023dlstack} characterized dependency bugs in the DL stack, identifying root causes, lifecycle stages, and fix patterns. 
These challenges are further exacerbated in PTM supply chains. Zhao \emph{et al.}~\cite{zhao2024malhug} presented a systematic study of malicious code poisoning attacks on pre-trained model hubs like Hugging Face, proposing a pipeline to detect and mitigate such threats. Similarly, Casey \emph{et al.}~\cite{casey2024mlattack} investigated unsafe serialization methods in Hugging Face models, emphasizing the pervasiveness of exploitable vulnerabilities.
These findings highlight the critical need for improved security practices, vulnerability tracking, and automated mitigation strategies in both traditional and emerging software supply chains.  
Our work extends this line of research by analyzing the security risks specific to LLM supply chains.

%% file: Chapters/8.conclusion.tex
\section{Conclusion}
\label{sec:conclusion}
This paper presents the first empirical study of the LLMSC, analyzing 13,486 open-source packages and 180 vulnerabilities. Our findings reveal that the LLMSC has a shallow, sparse hierarchy with a ``locally dense, globally sparse'' topology dominated by core foundational libraries, which act as critical hubs but pose significant security risks. We also identify dynamic growth in functional domains and persistent vulnerabilities that propagate through dependency chains, amplifying risks. This study provides a foundational understanding of the LLMSC and highlights the need for further research on securing this evolving ecosystem.

\section*{Acknowledgment}
This work was supported in part by the National Natural Science Foundation of China (grants No.62572209, 62502168) and the Hubei Provincial Key Research and Development Program (grant No. 2025BAB057).

%% file: reference.bib
@inproceedings{daye2024llmcodeunderstanding,
author = {Nam, Daye and Macvean, Andrew and Hellendoorn, Vincent and Vasilescu, Bogdan and Myers, Brad},
title = {Using an LLM to Help With Code Understanding},
year = {2024},
isbn = {9798400702174},
publisher = {Association for Computing Machinery},
address = {New York, NY, USA},
url = {https://doi.org/10.1145/3597503.3639187},
doi = {10.1145/3597503.3639187},
booktitle = {Proceedings of the IEEE/ACM 46th International Conference on Software Engineering},
articleno = {97},
numpages = {13},
location = {Lisbon, Portugal},
series = {ICSE '24}
}

@article{lu2024visionunderstanding,
  author       = {Haoyu Lu and
                  Wen Liu and
                  Bo Zhang and
                  Bingxuan Wang and
                  Kai Dong and
                  Bo Liu and
                  Jingxiang Sun and
                  Tongzheng Ren and
                  Zhuoshu Li and
                  Hao Yang and
                  Yaofeng Sun and
                  Chengqi Deng and
                  Hanwei Xu and
                  Zhenda Xie and
                  Chong Ruan},
  title        = {DeepSeek-VL: Towards Real-World Vision-Language Understanding},
  journal      = {CoRR},
  volume       = {abs/2403.05525},
  year         = {2024},
  url          = {https://doi.org/10.48550/arXiv.2403.05525},
  doi          = {10.48550/ARXIV.2403.05525},
  eprinttype    = {arXiv},
  eprint       = {2403.05525},
  bibsource    = {dblp computer science bibliography, https://dblp.org}
}

@article{zhao2024llmsurvey,
  author       = {Wayne Xin Zhao and
                  Kun Zhou and
                  Junyi Li and
                  Tianyi Tang and
                  Xiaolei Wang and
                  Yupeng Hou and
                  Yingqian Min and
                  Beichen Zhang and
                  Junjie Zhang and
                  Zican Dong and
                  Yifan Du and
                  Chen Yang and
                  Yushuo Chen and
                  Zhipeng Chen and
                  Jinhao Jiang and
                  Ruiyang Ren and
                  Yifan Li and
                  Xinyu Tang and
                  Zikang Liu and
                  Peiyu Liu and
                  Jian{-}Yun Nie and
                  Ji{-}Rong Wen},
  title        = {A Survey of Large Language Models},
  journal      = {CoRR},
  volume       = {abs/2303.18223},
  year         = {2023},
  url          = {https://doi.org/10.48550/arXiv.2303.18223},
  doi          = {10.48550/ARXIV.2303.18223},
  eprinttype    = {arXiv},
  eprint       = {2303.18223},
  bibsource    = {dblp computer science bibliography, https://dblp.org}
}

@article{roberto2023generativeai,
  author       = {Roberto Gozalo{-}Brizuela and
                  Eduardo C. Garrido{-}Merch{\'{a}}n},
  title        = {ChatGPT is not all you need. {A} State of the Art Review of large
                  Generative {AI} models},
  journal      = {CoRR},
  volume       = {abs/2301.04655},
  year         = {2023},
  url          = {https://doi.org/10.48550/arXiv.2301.04655},
  doi          = {10.48550/ARXIV.2301.04655},
  eprinttype    = {arXiv},
  eprint       = {2301.04655},
  bibsource    = {dblp computer science bibliography, https://dblp.org}
}

@article{hou2024llm4se,
author = {Hou, Xinyi and Zhao, Yanjie and Liu, Yue and Yang, Zhou and Wang, Kailong and Li, Li and Luo, Xiapu and Lo, David and Grundy, John and Wang, Haoyu},
title = {Large Language Models for Software Engineering: A Systematic Literature Review},
year = {2024},
issue_date = {November 2024},
publisher = {Association for Computing Machinery},
address = {New York, NY, USA},
volume = {33},
number = {8},
issn = {1049-331X},
url = {https://doi.org/10.1145/3695988},
doi = {10.1145/3695988},
journal = {ACM Trans. Softw. Eng. Methodol.},
month = dec,
articleno = {220},
numpages = {79}
}

@article{jin2024agents4se,
  author       = {Haolin Jin and
                  Linghan Huang and
                  Haipeng Cai and
                  Jun Yan and
                  Bo Li and
                  Huaming Chen},
  title        = {From LLMs to LLM-based Agents for Software Engineering: {A} Survey
                  of Current, Challenges and Future},
  journal      = {CoRR},
  volume       = {abs/2408.02479},
  year         = {2024},
  url          = {https://doi.org/10.48550/arXiv.2408.02479},
  doi          = {10.48550/ARXIV.2408.02479},
  eprinttype    = {arXiv},
  eprint       = {2408.02479},
  bibsource    = {dblp computer science bibliography, https://dblp.org}
}

@article{wang2024agentsinse,
  author       = {Yanlin Wang and
                  Wanjun Zhong and
                  Yanxian Huang and
                  Ensheng Shi and
                  Min Yang and
                  Jiachi Chen and
                  Hui Li and
                  Yuchi Ma and
                  Qianxiang Wang and
                  Zibin Zheng},
  title        = {Agents in Software Engineering: Survey, Landscape, and Vision},
  journal      = {CoRR},
  volume       = {abs/2409.09030},
  year         = {2024},
  url          = {https://doi.org/10.48550/arXiv.2409.09030},
  doi          = {10.48550/ARXIV.2409.09030},
  eprinttype    = {arXiv},
  eprint       = {2409.09030},
  bibsource    = {dblp computer science bibliography, https://dblp.org}
}

@article{wu2023autogen,
  author       = {Qingyun Wu and
                  Gagan Bansal and
                  Jieyu Zhang and
                  Yiran Wu and
                  Shaokun Zhang and
                  Erkang Zhu and
                  Beibin Li and
                  Li Jiang and
                  Xiaoyun Zhang and
                  Chi Wang},
  title        = {AutoGen: Enabling Next-Gen {LLM} Applications via Multi-Agent Conversation
                  Framework},
  journal      = {CoRR},
  volume       = {abs/2308.08155},
  year         = {2023},
  url          = {https://doi.org/10.48550/arXiv.2308.08155},
  doi          = {10.48550/ARXIV.2308.08155},
  eprinttype    = {arXiv},
  eprint       = {2308.08155},
  bibsource    = {dblp computer science bibliography, https://dblp.org}
}

@article{wang2024ala,
  author       = {Lei Wang and
                  Chen Ma and
                  Xueyang Feng and
                  Zeyu Zhang and
                  Hao Yang and
                  Jingsen Zhang and
                  Zhiyuan Chen and
                  Jiakai Tang and
                  Xu Chen and
                  Yankai Lin and
                  Wayne Xin Zhao and
                  Zhewei Wei and
                  Jirong Wen},
  title        = {A survey on large language model based autonomous agents},
  journal      = {Frontiers Comput. Sci.},
  volume       = {18},
  number       = {6},
  pages        = {186345},
  year         = {2024},
  url          = {https://doi.org/10.1007/s11704-024-40231-1},
  doi          = {10.1007/S11704-024-40231-1},
  bibsource    = {dblp computer science bibliography, https://dblp.org}
}

@misc{owasp,
author = {OWASP},
title = {OWASP Top 10 for Large Language Model Applications},
year = {2025},
howpublished = {\url{https://owasp.org/www-project-top-10-for-large-language-model-applications/}},
note = {Accessed: 2026-01-12}
}

@article{wang2024llmsc,
author = {Wang, Shenao and Zhao, Yanjie and Hou, Xinyi and Wang, Haoyu},
title = {Large Language Model Supply Chain: A Research Agenda},
year = {2024},
publisher = {Association for Computing Machinery},
address = {New York, NY, USA},
issn = {1049-331X},
url = {https://doi.org/10.1145/3708531},
doi = {10.1145/3708531},
note = {Just Accepted},
journal = {ACM Trans. Softw. Eng. Methodol.},
month = dec
}

@article{hu2024llmsc,
  author       = {Qiang Hu and
                  Xiaofei Xie and
                  Sen Chen and
                  Lei Ma},
  title        = {Large Language Model Supply Chain: Open Problems From the Security
                  Perspective},
  journal      = {CoRR},
  volume       = {abs/2411.01604},
  year         = {2024},
  url          = {https://doi.org/10.48550/arXiv.2411.01604},
  doi          = {10.48550/ARXIV.2411.01604},
  eprinttype    = {arXiv},
  eprint       = {2411.01604},
  bibsource    = {dblp computer science bibliography, https://dblp.org}
}

@article{huang2024llmsc,
  author       = {Kaifeng Huang and
                  Bihuan Chen and
                  You Lu and
                  Susheng Wu and
                  Dingji Wang and
                  Yiheng Huang and
                  Haowen Jiang and
                  Zhuotong Zhou and
                  Junming Cao and
                  Xin Peng},
  title        = {Lifting the Veil on the Large Language Model Supply Chain: Composition,
                  Risks, and Mitigations},
  journal      = {CoRR},
  volume       = {abs/2410.21218},
  year         = {2024},
  url          = {https://doi.org/10.48550/arXiv.2410.21218},
  doi          = {10.48550/ARXIV.2410.21218},
  eprinttype    = {arXiv},
  eprint       = {2410.21218},
  bibsource    = {dblp computer science bibliography, https://dblp.org}
}

@inproceedings{zhao2024malhug,
author = {Zhao, Jian and Wang, Shenao and Zhao, Yanjie and Hou, Xinyi and Wang, Kailong and Gao, Peiming and Zhang, Yuanchao and Wei, Chen and Wang, Haoyu},
title = {Models Are Codes: Towards Measuring Malicious Code Poisoning Attacks on Pre-trained Model Hubs},
year = {2024},
isbn = {9798400712487},
publisher = {Association for Computing Machinery},
address = {New York, NY, USA},
url = {https://doi.org/10.1145/3691620.3695271},
doi = {10.1145/3691620.3695271},
booktitle = {Proceedings of the 39th IEEE/ACM International Conference on Automated Software Engineering},
pages = {2087–2098},
numpages = {12},
location = {Sacramento, CA, USA},
series = {ASE '24}
}

@inproceedings{jiang2022ptm,
author = {Jiang, Wenxin and Synovic, Nicholas and Sethi, Rohan and Indarapu, Aryan and Hyatt, Matt and Schorlemmer, Taylor R. and Thiruvathukal, George K. and Davis, James C.},
title = {An Empirical Study of Artifacts and Security Risks in the Pre-trained Model Supply Chain},
year = {2022},
isbn = {9781450398855},
publisher = {Association for Computing Machinery},
address = {New York, NY, USA},
url = {https://doi.org/10.1145/3560835.3564547},
doi = {10.1145/3560835.3564547},
booktitle = {Proceedings of the 2022 ACM Workshop on Software Supply Chain Offensive Research and Ecosystem Defenses},
pages = {105–114},
numpages = {10},
location = {Los Angeles, CA, USA},
series = {SCORED'22}
}

@inproceedings{chen2024agentpoison,
  author       = {Zhaorun Chen and
                  Zhen Xiang and
                  Chaowei Xiao and
                  Dawn Song and
                  Bo Li},
  editor       = {Amir Globersons and
                  Lester Mackey and
                  Danielle Belgrave and
                  Angela Fan and
                  Ulrich Paquet and
                  Jakub M. Tomczak and
                  Cheng Zhang},
  title        = {AgentPoison: Red-teaming {LLM} Agents via Poisoning Memory or Knowledge
                  Bases},
  booktitle    = {Advances in Neural Information Processing Systems 38: Annual Conference
                  on Neural Information Processing Systems 2024, NeurIPS 2024, Vancouver,
                  BC, Canada, December 10 - 15, 2024},
  year         = {2024},
  url          = {http://papers.nips.cc/paper\_files/paper/2024/hash/eb113910e9c3f6242541c1652e30dfd6-Abstract-Conference.html},
  bibsource    = {dblp computer science bibliography, https://dblp.org}
}

@inproceedings{zhang2024ragpoisoning,
author = {Zhang, Quan and Zeng, Binqi and Zhou, Chijin and Go, Gwihwan and Shi, Heyuan and Jiang, Yu},
title = {Human-Imperceptible Retrieval Poisoning Attacks in LLM-Powered Applications},
year = {2024},
isbn = {9798400706585},
publisher = {Association for Computing Machinery},
address = {New York, NY, USA},
url = {https://doi.org/10.1145/3663529.3663786},
doi = {10.1145/3663529.3663786},
booktitle = {Companion Proceedings of the 32nd ACM International Conference on the Foundations of Software Engineering},
pages = {502–506},
numpages = {5},
location = {Porto de Galinhas, Brazil},
series = {FSE 2024}
}

@INPROCEEDINGS{jiang2023huggingface,
  author={Jiang, Wenxin and Synovic, Nicholas and Hyatt, Matt and Schorlemmer, Taylor R. and Sethi, Rohan and Lu, Yung-Hsiang and Thiruvathukal, George K. and Davis, James C.},
  booktitle={2023 IEEE/ACM 45th International Conference on Software Engineering (ICSE)}, 
  title={An Empirical Study of Pre-Trained Model Reuse in the Hugging Face Deep Learning Model Registry}, 
  year={2023},
  volume={},
  number={},
  pages={2463-2475},
  doi={10.1109/ICSE48619.2023.00206}
}

@misc{llamaindexcve,
author = {Github},
title = {llama-index vulnerable to arbitrary code execution},
year = {2024},
howpublished = {\url{https://github.com/advisories/GHSA-2xxc-73fv-36f7}},
note = {Accessed: 2026-01-12}
}

@misc{transformerscve,
author = {Github},
title = {transformers has a Deserialization of Untrusted Data vulnerability},
year = {2024},
howpublished = {\url{https://github.com/advisories/GHSA-3863-2447-669p}},
note = {Accessed: 2026-01-12}
}

@inproceedings{tan2022dlsc,
author = {Tan, Xin and Gao, Kai and Zhou, Minghui and Zhang, Li},
title = {An exploratory study of deep learning supply chain},
year = {2022},
isbn = {9781450392211},
publisher = {Association for Computing Machinery},
address = {New York, NY, USA},
url = {https://doi.org/10.1145/3510003.3510199},
doi = {10.1145/3510003.3510199},
booktitle = {Proceedings of the 44th International Conference on Software Engineering},
pages = {86–98},
numpages = {13},
location = {Pittsburgh, Pennsylvania},
series = {ICSE '22}
}

@article{gao2024dlsc,
author = {Gao, Kai and He, Runzhi and Xie, Bing and Zhou, Minghui},
title = {Characterizing Deep Learning Package Supply Chains in PyPI: Domains, Clusters, and Disengagement},
year = {2024},
issue_date = {May 2024},
publisher = {Association for Computing Machinery},
address = {New York, NY, USA},
volume = {33},
number = {4},
issn = {1049-331X},
url = {https://doi.org/10.1145/3640336},
doi = {10.1145/3640336},
journal = {ACM Trans. Softw. Eng. Methodol.},
month = apr,
articleno = {97},
numpages = {27}
}

@misc{tensorflow,
author = {tensorflow},
title = {tensorflow},
year = {2025},
howpublished = {\url{https://github.com/tensorflow/tensorflow}},
note = {Accessed: 2026-01-12}
}

@misc{pytorch,
author = {pytorch},
title = {pytorch},
year = {2025},
howpublished = {\url{https://github.com/pytorch/pytorch}},
note = {Accessed: 2026-01-12}
}

@inproceedings{liu2022npmsc,
author = {Liu, Chengwei and Chen, Sen and Fan, Lingling and Chen, Bihuan and Liu, Yang and Peng, Xin},
title = {Demystifying the vulnerability propagation and its evolution via dependency trees in the NPM ecosystem},
year = {2022},
isbn = {9781450392211},
publisher = {Association for Computing Machinery},
address = {New York, NY, USA},
url = {https://doi.org/10.1145/3510003.3510142},
doi = {10.1145/3510003.3510142},
booktitle = {Proceedings of the 44th International Conference on Software Engineering},
pages = {672–684},
numpages = {13},
location = {Pittsburgh, Pennsylvania},
series = {ICSE '22}
}

@INPROCEEDINGS{wu2023mavensc,
  author={Wu, Yulun and Yu, Zeliang and Wen, Ming and Li, Qiang and Zou, Deqing and Jin, Hai},
  booktitle={2023 IEEE/ACM 45th International Conference on Software Engineering (ICSE)}, 
  title={Understanding the Threats of Upstream Vulnerabilities to Downstream Projects in the Maven Ecosystem}, 
  year={2023},
  volume={},
  number={},
  pages={1046-1058},
  doi={10.1109/ICSE48619.2023.00095}}

@inproceedings{tang2023csc,
author = {Tang, Wei and Xu, Zhengzi and Liu, Chengwei and Wu, Jiahui and Yang, Shouguo and Li, Yi and Luo, Ping and Liu, Yang},
title = {Towards Understanding Third-party Library Dependency in C/C++ Ecosystem},
year = {2023},
isbn = {9781450394758},
publisher = {Association for Computing Machinery},
address = {New York, NY, USA},
url = {https://doi.org/10.1145/3551349.3560432},
doi = {10.1145/3551349.3560432},
booktitle = {Proceedings of the 37th IEEE/ACM International Conference on Automated Software Engineering},
articleno = {106},
numpages = {12},
location = {Rochester, MI, USA},
series = {ASE '22}
}

@article{wang2024agentsurvey,
  author       = {Lei Wang and
                  Chen Ma and
                  Xueyang Feng and
                  Zeyu Zhang and
                  Hao Yang and
                  Jingsen Zhang and
                  Zhiyuan Chen and
                  Jiakai Tang and
                  Xu Chen and
                  Yankai Lin and
                  Wayne Xin Zhao and
                  Zhewei Wei and
                  Jirong Wen},
  title        = {A survey on large language model based autonomous agents},
  journal      = {Frontiers Comput. Sci.},
  volume       = {18},
  number       = {6},
  pages        = {186345},
  year         = {2024},
  url          = {https://doi.org/10.1007/s11704-024-40231-1},
  doi          = {10.1007/S11704-024-40231-1},
  bibsource    = {dblp computer science bibliography, https://dblp.org}
}

@article{hadi2025large,
  title={Large Language Models: A Comprehensive Survey of its Applications, Challenges, Limitations, and Future Prospects},
  author={Hadi, Muhammad Usman and Al Tashi, Qasem and Qureshi, Rizwan and others},
  journal={TechRxiv},
  year={2025},
  month={February},
  day={10},
  doi={10.36227/techrxiv.23589741.v8},
  url={https://doi.org/10.36227/techrxiv.23589741.v8}
}

@INPROCEEDINGS{diaz2024llmops,
  author={Diaz-De-Arcaya, Josu and López-De-Armentia, Juan and Miñón, Raúl and Ojanguren, Iker Lasa and Torre-Bastida, Ana I.},
  booktitle={2024 9th International Conference on Smart and Sustainable Technologies (SpliTech)}, 
  title={Large Language Model Operations (LLMOps): Definition, Challenges, and Lifecycle Management}, 
  year={2024},
  volume={},
  number={},
  pages={1-4},
  doi={10.23919/SpliTech61897.2024.10612341}}

@article{wang2025sok,
  author       = {Shenao Wang and
                  Yanjie Zhao and
                  Zhao Liu and
                  Quanchen Zou and
                  Haoyu Wang},
  title        = {SoK: Understanding Vulnerabilities in the Large Language Model Supply
                  Chain},
  journal      = {CoRR},
  volume       = {abs/2502.12497},
  year         = {2025},
  url          = {https://doi.org/10.48550/arXiv.2502.12497},
  doi          = {10.48550/ARXIV.2502.12497},
  eprinttype    = {arXiv},
  eprint       = {2502.12497},
  bibsource    = {dblp computer science bibliography, https://dblp.org}
}

@misc{pypi,
author = {PyPI},
title = {PyPI},
year = {2025},
howpublished = {\url{https://pypi.org/}},
note = {Accessed: 2026-01-12}
}

@misc{npm,
author = {NPM},
title = {Registry},
year = {2025},
howpublished = {\url{https://github.com/npm/registry}},
note = {Accessed: 2026-01-12}
}

@misc{deepseek,
author = {deepseek-ai},
title = {DeepSeek-V3},
year = {2025},
howpublished = {\url{https://github.com/deepseek-ai/DeepSeek-V3}},
note = {Accessed: 2026-01-12}
}

@misc{nvd,
author = {NVD},
title = {CVE Search},
year = {2025},
howpublished = {\url{https://nvd.nist.gov/vuln/search}},
note = {Accessed: 2026-01-12}
}

@misc{gitadvisory,
author = {Github},
title = {Github Advisory Database},
year = {2025},
howpublished = {\url{https://github.com/advisories}},
note = {Accessed: 2026-01-12}
}

@misc{huntr,
author = {Huntr},
title = {The world’s first bug bounty platform for AI/ML},
year = {2025},
howpublished = {\url{https://huntr.com/}},
note = {Accessed: 2026-01-12}
}

@misc{librariesio,
author = {Libraries.io},
title = {Libraries.io: security \& maintenance data for open source software},
year = {2025},
howpublished = {\url{https://libraries.io/}},
note = {Accessed: 2026-01-12}
}

@inproceedings{ma2024web3sc,
  author       = {Kai Ma and
                  Zhuo Wang and
                  Yanjie Zhao and
                  Haoyu Wang},
  editor       = {Hong Mei and
                  Jian Lv and
                  Abdelsalam Helal and
                  Xiaoxing Ma and
                  Shing{-}Chi Cheung and
                  Jie Zhang and
                  Tao Zhang},
  title        = {Decoding Web3: In-depth Analysis of the Third-Party Package Supply
                  Chain},
  booktitle    = {Proceedings of the 15th Asia-Pacific Symposium on Internetware, Internetware
                  2024, Macau, SAR, China, July 24-26, 2024},
  publisher    = {{ACM}},
  year         = {2024},
  url          = {https://doi.org/10.1145/3671016.3671402},
  doi          = {10.1145/3671016.3671402},
  bibsource    = {dblp computer science bibliography, https://dblp.org}
}

@INPROCEEDINGS{mir2023propagation,
  author={Mir, Amir M. and Keshani, Mehdi and Proksch, Sebastian},
  booktitle={2023 IEEE International Conference on Software Analysis, Evolution and Reengineering (SANER)}, 
  title={On the Effect of Transitivity and Granularity on Vulnerability Propagation in the Maven Ecosystem}, 
  year={2023},
  volume={},
  number={},
  pages={201-211},
  doi={10.1109/SANER56733.2023.00028}}

@inproceedings{wu2024vulpropa,
author = {Wu, Susheng and Wang, Ruisi and Huang, Kaifeng and Cao, Yiheng and Song, Wenyan and Zhou, Zhuotong and Huang, Yiheng and Chen, Bihuan and Peng, Xin},
title = {Vision: Identifying Affected Library Versions for Open Source Software Vulnerabilities},
year = {2024},
isbn = {9798400712487},
publisher = {Association for Computing Machinery},
address = {New York, NY, USA},
url = {https://doi.org/10.1145/3691620.3695516},
doi = {10.1145/3691620.3695516},
booktitle = {Proceedings of the 39th IEEE/ACM International Conference on Automated Software Engineering},
pages = {1447–1459},
numpages = {13},
location = {Sacramento, CA, USA},
series = {ASE '24}
}

@INPROCEEDINGS{zhang2023maven,
  author={Zhang, Lyuye and Liu, Chengwei and Chen, Sen and Xu, Zhengzi and Fan, Lingling and Zhao, Lida and Zhang, Yiran and Liu, Yang},
  booktitle={2023 38th IEEE/ACM International Conference on Automated Software Engineering (ASE)}, 
  title={Mitigating Persistence of Open-Source Vulnerabilities in Maven Ecosystem}, 
  year={2023},
  volume={},
  number={},
  pages={191-203},
  doi={10.1109/ASE56229.2023.00058}}

@inproceedings{jiang2024peatmoss,
author = {Jiang, Wenxin and Yasmin, Jerin and Jones, Jason and Synovic, Nicholas and Kuo, Jiashen and Bielanski, Nathaniel and Tian, Yuan and Thiruvathukal, George K. and Davis, James C.},
title = {PeaTMOSS: A Dataset and Initial Analysis of Pre-Trained Models in Open-Source Software},
year = {2024},
isbn = {9798400705878},
publisher = {Association for Computing Machinery},
address = {New York, NY, USA},
url = {https://doi.org/10.1145/3643991.3644907},
doi = {10.1145/3643991.3644907},
booktitle = {Proceedings of the 21st International Conference on Mining Software Repositories},
pages = {431–443},
numpages = {13},
location = {Lisbon, Portugal},
series = {MSR '24}
}

@article{stalnaker2025mlsc,
  author       = {Trevor Stalnaker and
                  Nathan Wintersgill and
                  Oscar Chaparro and
                  Laura A. Heymann and
                  Massimiliano Di Penta and
                  Daniel M. Germ{\'{a}}n and
                  Denys Poshyvanyk},
  title        = {The {ML} Supply Chain in the Era of Software 2.0: Lessons Learned
                  from Hugging Face},
  journal      = {CoRR},
  volume       = {abs/2502.04484},
  year         = {2025},
  url          = {https://doi.org/10.48550/arXiv.2502.04484},
  doi          = {10.48550/ARXIV.2502.04484},
  eprinttype    = {arXiv},
  eprint       = {2502.04484},
  bibsource    = {dblp computer science bibliography, https://dblp.org}
}

@article{alfadel2023pypi,
  author       = {Mahmoud Alfadel and
                  Diego Elias Costa and
                  Emad Shihab},
  title        = {Empirical analysis of security vulnerabilities in Python packages},
  journal      = {Empir. Softw. Eng.},
  volume       = {28},
  number       = {3},
  pages        = {59},
  year         = {2023},
  url          = {https://doi.org/10.1007/s10664-022-10278-4},
  doi          = {10.1007/S10664-022-10278-4},
  bibsource    = {dblp computer science bibliography, https://dblp.org}
}

@inproceedings{huang2023dlstack,
author = {Huang, Kaifeng and Chen, Bihuan and Wu, Susheng and Cao, Junming and Ma, Lei and Peng, Xin},
title = {Demystifying Dependency Bugs in Deep Learning Stack},
year = {2023},
isbn = {9798400703270},
publisher = {Association for Computing Machinery},
address = {New York, NY, USA},
url = {https://doi.org/10.1145/3611643.3616325},
doi = {10.1145/3611643.3616325},
booktitle = {Proceedings of the 31st ACM Joint European Software Engineering Conference and Symposium on the Foundations of Software Engineering},
pages = {450–462},
numpages = {13},
location = {San Francisco, CA, USA},
series = {ESEC/FSE 2023}
}

@article{casey2024mlattack,
  author       = {Beatrice Casey and
                  Joanna C. S. Santos and
                  Mehdi Mirakhorli},
  title        = {A Large-Scale Exploit Instrumentation Study of {AI/ML} Supply Chain
                  Attacks in Hugging Face Models},
  journal      = {CoRR},
  volume       = {abs/2410.04490},
  year         = {2024},
  url          = {https://doi.org/10.48550/arXiv.2410.04490},
  doi          = {10.48550/ARXIV.2410.04490},
  eprinttype    = {arXiv},
  eprint       = {2410.04490},
  bibsource    = {dblp computer science bibliography, https://dblp.org}
}
